\documentclass[aps,prc,twocolumn,groupedaddress]{revtex4}
\usepackage{amsmath}
\usepackage{graphicx}
\usepackage{subfigure}

\begin{document}
\title{Combinatorial nuclear level-density model}
\author{H. Uhrenholt$^1$, S. {\AA}berg$^1$, P. M\"{o}ller$^2$, T. Ichikawa$^3$}
\affiliation{$^1$Division of Mathematical Physics, LTH, Lund University,
  P.O. Box 118, S-221 00 Lund, Sweden\\
$^2$Theoretical Division, Los Alamos National Laboratory, Los Alamos, NM
87545, USA\\
$^3$ RIKEN Nishina Center, RIKEN, Wako, Saitama 351-0198, Japan}
\begin{abstract}
A microscopic nuclear level-density model is presented. The model is
a completely combinatorial (micro-canonical) model based on the
folded-Yukawa single-particle potential and includes explicit
treatment of pairing, rotational and vibrational states. The
microscopic character of all states enables extraction of level
distribution functions with respect to pairing gaps, parity and
angular momentum. The results of the model are compared to available
experimental data: neutron separation energy level spacings, data on
total level-density functions from the Oslo method, cumulative level
densities from low-lying discrete states, and data on parity ratios.
\end{abstract}
\maketitle

%%%%%%%%%%%%%%%%%%%%%%%%%%%%%%%%%%%%%%%%%%%%%%%%%%%%%%%%%%%%%%%

\section{Introduction}

Nuclear many-body level-density models are  key ingredients in
nuclear reaction theories, where they, for example, govern the rates
and decay patterns of  astrophysical processes and nuclear fission.
In statistical methods, for example the Hauser-Feshbach formalism
\cite{HauserFeshbach} for describing nuclear reactions, a knowledge
of the level density is crucial
\cite{RauscherAstro,MollerFission,Rauscher}. How to calculate the
nuclear level density (NLD) has been a long-standing challenge
\cite{Bethe,Ericson,Bjornholm,BMII,Soloviev,GilbertCameron}.
Recently it has been subject
to renewed interest, theoretically as well as experimentally.%, see eg. Refs.~\cite{OsloMethod,RichterParity,Goriely2006}.

The simplest type of model is the Fermi-gas model, which is based on
the partition-function method. It provides simple analytical
formulas for the NLD \cite{BMI}. Several phenomenological extensions
have been proposed in order to reproduce experimental data. By
adjusting free model parameters to data these models give
unprecedented accuracy in the region of the parameter fit
\cite{Rauscher,GSM}. Also semi-classical methods have been used to
obtain expressions for the level density \cite{Leboeuf}. However,
the Fermi-gas models are unreliable outside these regions, eg. when
they are extrapolated to higher excitation energies or to nucleon
numbers far from stability. Ideally nuclear structure should be
included in NLD models. Several combinatorial models based on
nuclear mean-field theory have been proposed, see eg.
Refs.~\cite{Goriely2006,HilaireGirod}. Beyond mean-field methods
have also been used to model the NLD, eg. the Shell-Model
Monte-Carlo method \cite{AlhassidParity,AlhassidSpinDist} and the
interacting shell model \cite{Horoi}. These latter models take into
account effective nucleon-nucleon interactions, but at the same time
suffer from limitations due to the limited size of the Hilbert
space, and hence are presently unable to provide global predictions
for level spacings at the neutron-separation energy.

Experimentally the NLD has been subject to renewed interest in the
last decade partly due to the development of the Oslo method, which
has provided  new experimental data \cite{OsloMethod}. The Oslo
method provides  the level density over extended regions of
excitation energy as opposed to the neutron-separation
level-spacings data which only provide one data point at relatively
high excitation energy. Also recent measurements of separate level
densities of $2^+$ and $2^-$ states \cite{RichterParity} challenge
theory  to reproduce these observed parity ratios.

Few nuclear-structure models have been used to simultaneously
globally describe nuclear masses, fission barriers, ground-state
spins and decay rates. One such model is the microscopic-macroscopic
FRLDM model which has previously been used to model these
observables
\cite{MollerFission,MollerMass,GroundStateSpins,MollerASTRO}, and
here serve as the starting point for calculating the nuclear level
density. In this paper a combinatorial (micro-canonical) nuclear
level-density model based on the folded-Yukawa single-particle model
is presented. The model is fully microscopic with pairing
correlations, vibrations, and rotational excitations calculated for
each many-particle-many-hole excited state. Presently, the lowest
ten million or so states can be accounted for. This implies that the
excitation energy region from the ground state to well above the neutron
resonance region is included. No additional
parameters are introduced in the model, and no refitting of
parameters of the FRLDM is performed.

In Sec.~\ref{sec:nld} the combinatorial folded-Yukawa (CFY)
level-density model is described. The model allows explicit tracking
of quantum numbers, and distributions of pairing gaps, parity and
angular momentum are discussed in Sec.~\ref{sec:leveldistr}. Results
from the CFY model are compared to experimental data in
Sec.~\ref{sec:EXP}, and in Sec.~\ref{sec:othermodels} the CFY model
is compared to other theoretical NLD models. Finally, a short
summary is given in Sec.~\ref{sec:summary}.

%%%%%%%%%%%%%%%%%%%%%%%%%%%%%%%%%%%%%%%%%%%%%%%%%%%%%%%%%%%%%%%

\section{The combinatorial model of NLD (CFY)}
\label{sec:nld}

The NLD is calculated by means of a combinatorial counting of
excited many-particle-many-hole states as described in
Sec.~\ref{sec:comb}. In Sec.~\ref{sec:pairing} we present how
pairing is taken into account for excited states by explicitly
solving the BCS equations for all individual configurations.
Rotations are taken into account combinatorially with a
pairing-dependent moment of inertia (Sec.~\ref{sec:ROT}). The
vibrational contribution to the NLD is investigated by including
microscopically described phonons using the Quasi-particle
Tamm-Dancoff Approximation (QTDA), see Sec.~\ref{sec:VIB}. In
Sec.~\ref{sec:spread} it is described how we briefly account for a
general residual interaction causing a smearing of the level-density
distribution at higher excitation energies.

All produced nuclear levels, calculated as described below, are
sorted into a binned level density, where the typical bin size is in
the range $\Delta E$ = 30--50 keV\@. The level density is calculated
by counting the number of levels in the energy bin,
\begin{equation}
  \rho(E_b,I,\pi) = \frac{1}{\Delta E}\int_{E_b-\frac{\Delta E}{2}}^{E_b+\frac{\Delta E}{2}}
  \sum_i \delta (E-E_i(I,\pi)) dE,
  \label{levdendef}
\end{equation}
where $E_b$ is the energy center of bin $b$ and $E_i(I,\pi)$ denotes
the calculated state with energy $E_i$ (given by
Eq.~(\ref{TotalEnergy})), angular momentum $I$ and parity $\pi$. The
total level density at a given excitation energy $E$ is
\begin{equation}
  \rho_{\rm tot}(E) = \sum_{I,\pi} \rho(E,I,\pi).
\end{equation}

\subsection{Combinatorial intrinsic level density}
\label{sec:comb}

A high-quality combinatorial level-density model requires a realistic
description of the single-particle energies. We shall here utilize a
well-tested mean-field model, namely the microscopic-macroscopic
finite range liquid drop model (FRLDM) \cite{MollerMass}, that
provides a good global description of several nuclear-structure
properties such as masses, fission barriers and beta-decay
properties. The good agreement between  ground-state spins
calculated in the FRLDM model and experiment implies that the
single-particle spectrum close to the Fermi surface is well
described \cite{GroundStateSpins}.

The single-particle energies are thus obtained by solving the
one-body Schr\"{o}dinger equation,
\begin{equation}
(T+V_{FY}(\bar{\varepsilon})) | \nu \rangle = e_{\nu} | \nu \rangle
,
\end{equation}
for protons as well as for neutrons, where $V_{FY}$ is the
Folded-Yukawa single-particle potential. The parameters of the
potential, as well as the deformation parameters,
$\bar{\varepsilon}=(\varepsilon_2,\varepsilon_3,\varepsilon_4,
...)$, are taken from an extensive calculation of nuclear masses
\cite{MollerMass}. All parameters of the single-particle model are
thus fixed from other studies. One aim of the present study is to
see how well highly excited states (in terms of level-densities) are
described for all nuclei heavier than $^{16}$O.

The many-body ground state (the many-body particle vacuum) $| 0
\rangle$, is obtained by filling the lowest $N$ ($Z$) states
(including time-reversed states),
\begin{equation}
|0 \rangle = \sum_{\nu=1}^{N \mbox{ } (Z)} a_{\nu}^+ | - \rangle,
\end{equation}
where $| - \rangle $ is the one-body vacuum. Intrinsic excitations
are obtained by many-particle-many-hole excitations on the many-body
particle vacuum,
\begin{equation}
|i \rangle = \prod_{\alpha = 1}^n a_{\nu_{\alpha}}^ +
a_{\nu'_{\alpha}} | 0 \rangle,
\label{many-body_wf}
\end{equation}
where $\nu$ and $\nu'$ span all single-particle states (as well as
time-reversed states) in the potential. We include all
$n$-particle-$n$-hole states with $n<10$ for protons as well as for
neutrons. Corresponding energies are given by
\begin{equation}
E_i = E_0 + \sum_{\alpha=1}^n (e_{\nu_{\alpha}} - e_{\nu'_{\alpha}}
),
\end{equation}
where $E_0$ is the ground-state energy. (When pairing is included
this expression is modified as described in Sec.~\ref{sec:pairing}.)
Nuclei are assumed to have constant deformation for all excitation
energies. For spherical and well-deformed nuclei this assumption is
expected to be a good approximation for excitation energies here
considered (mainly below the neutron separation energy).
For transitional nuclei this approximation might, however, introduce
some inaccuracies in the calculations.

In addition to the energy and the number of unpaired nucleons
(seniority), we also register the total parity and $K$-quantum
number for each excited state, where $K$ as usual is the angular
momentum projection on an intrinsic symmetry axis. The
identification of a $K$-quantum number can be done for all nuclei by
assuming an infinitesimal deformation for spherical nuclei. The
$K$-quantum number of the individual unpaired nucleons are allowed
to couple in all different ways to build up the total $K$-quantum
number.

%----------------------------------------------------

\subsection{Pairing}
\label{sec:pairing}

The many-body wave-function of the excited states is approximated by
the BCS wave-function with excited quasi-particles
\begin{eqnarray}
  \left| \tau \right> = \prod_{\nu'' \in \tau_2}
  (-V_{\nu''}+U_{\nu''}a_{\nu''}^\dagger a_{\bar{\nu}''}^\dagger ) \times\nonumber\\
  \times \prod_{\nu' \in \tau_1} a_{\nu'}^\dagger
  \prod_{\nu \in \tau_0}
  (U_{\nu}+V_{\nu}a_{\nu}^\dagger a_{\bar{\nu}}^\dagger ) \left| 0 \right>,
\end{eqnarray}
where $\tau_2,\tau_1$ and $\tau_0$ denote the spaces of double,
single and zero quasi-particle excitations, respectively, used to
build the many-body state $i$ of Eq.~\ref{many-body_wf}. $U_\nu$ and
$V_\nu$ are the standard BCS vacancy and occupation factors and
$\left|0\right>$ is the particle vacuum, see e.g.
Ref.~\cite{DossingBCS}. For the excited pairs in the group $\tau_2$
the effect (compared to zero-quasi-particle states) is simply
\begin{equation}
  U_\nu\rightarrow -V_\nu,\qquad V_\nu\rightarrow U_\nu.
  \label{UVchange}
\end{equation}

The pairing gap $\Delta$ and the Fermi energy $\lambda$ are obtained by solving
the BCS-equations
\begin{align}
  \Delta &= G \left[\sum_{\nu \in \tau_0}U_\nu V_\nu - \sum_{\nu'' \in \tau_2}
  U_{\nu''} V_{\nu''}\right]\label{BCSEQ1},
\\
N &= 2\sum_{\nu \in \tau_0} V^2_\nu + \sum_{\nu' \in \tau_1} 1 + 2\sum_{\nu''
  \in \tau_2} U^2_{\nu''}
\label{BCSEQ2}
\end{align}
for each state.

The pairing strength in the BCS model is governed by a single,
primary BCS model parameter, namely the parameter of the
effective-interaction pairing gap ($r$ in Eq.\~(47) in
\cite{MollerPair} or the equivalent $c_{\rm p}$ in Eq. (49) and
$c_{\rm n}$ in Eq. (50) in \cite{OlofssonPairing}). In an adjustment
of macroscopic, pairing, and other model parameters to optimize a
nuclear mass model $r=4.8$ MeV was obtained for the standard BCS
pairing model \cite{MollerMass}.  Because the BCS model we use here,
in contrast to Ref.~\cite{MollerPair,MollerMass}, includes blocking
and particle-number-projection it is necessary to use a pairing
strength optimized for this method.  In Ref.\ \cite{OlofssonPairing}
$c_{\rm p} = c_{\rm n} = 4.95$ MeV were obtained in an adjustment of
calculated pairing gaps to odd-even mass differences.  However, it
has been pointed out \cite{MollerPair,MollerMass} that odd-even mass
differences are subject to numerous non-smooth contributions, for
example from deformation changes and from irregularities in the
microscopic level structure, not just odd-even staggering due to
pairing. It is therefore better to determine the pairing strength
from a full nuclear mass calculation that includes the particular
pairing model under consideration and the associated adjustment of
all model parameters \cite{MollerMass}.  In the folded-Yukawa
macroscopic-microscopic mass model this yields the value 4.5 MeV for
the optimized strength parameter of the particle-number-projected
BCS model \cite{OlofssonPairing}.  The pairing strength $G$, which
is used to calculate pairing gaps for all excited states in the
level-density calculation, is determined from this average pairing
gap for each nuclear system; for details see
Ref.~\cite{OlofssonPairing}.

It follows that the excitation energy of the intrinsic many-body
configuration for one nucleon type is
\begin{multline}
  E_{{\rm mb},t} = 2\sum_{\nu \in \tau_0} e_\nu V^2_\nu + \sum_{\nu' \in \tau_1}
  e_\nu + 2\sum_{\nu'' \in \tau_2} e_\nu U^2_{\nu''}-\\
  G\sum_{\nu \in \tau_0} V^4_\nu -\frac{G}{2} \sum_{\nu' \in \tau_1}
  1 - G\sum_{\nu'' \in \tau_2} U^4_{\nu''}
 -\frac{\Delta^2}{G} -E_t^0,
  \label{Emb}
\end{multline}
where $t$ denotes protons or neutrons and $E_t^0$ is the proton or
neutron part of the ground-state energy. The pairing gap and Fermi
level are calculated by solving the BCS equations
Eqs.~(\ref{BCSEQ1}) and (\ref{BCSEQ2}). The total intrinsic
excitation energy of a many-body state $i$ is
\begin{equation}
 E_{\rm  mb}^i=E_{\rm mb,p}+E_{\rm mb,n}.
 \label{Total}
\end{equation}

The combinatorial approach to calculate the level density involves
calculating all possible many-body states, $E_{\rm mb}^i$, which
thus means that the BCS equations are solved about 10$^7$ times for
each nucleus.

%----------------------------------------------------

\subsection{Rotations}
\label{sec:ROT}

A general feature of deformed nuclei is the existence of rotational
bands built on the ground state as well as on excited states. The
question is how to incorporate these states in the NLD. In
principle, rotational states may be microscopically treated by
solving the cranking Hamiltonian, $h^{\omega}=h^0-\omega_{rot}j_x$,
where $\omega_{rot}$ is the rotational frequency and $j_x$ is the
angular momentum operator for rotations around the $x$-axis. Matrix
elements of the $j_x$-operator correspond to energy excitations of
the order $\Delta E=\varepsilon_2 \hbar \omega \approx
\varepsilon_2\cdot 41A^{-1/3}$ MeV for a nucleus with mass number
$A$ and quadrupole deformation $\varepsilon_2$ \cite{NilssonModel}.
This energy can be compared to a typical energy of lowest rotational
state, $E_{2^+}=\frac{\hbar^2}{2\mathcal{J}}2(2+1)\approx 90
A^{-5/3}$ MeV, and we see that $\Delta E >> E_{2^+}$ for all
considered nuclei. Since the collective state and the corresponding
matrix element are well separated in energy, the risk of
double-counting states by adding a rotational band on a
microscopically calculated band-head is quite small, cf
Ref.~\cite{Bjornholm}. This is true as long as the excitation
energies are smaller than energies corresponding to the temperature,
$T=\Delta E$, i.e. for excitation energies smaller than about
$10A^{1/3}$ MeV (for $\varepsilon_2=0.25$), giving about 30 MeV for
mass A=25 and 55 MeV for A=160. For higher excitation energies a
saturation of the rotational enhancement should set in. In this
study we concentrate on lower energies, say up to about 10 MeV. It
is therefore a reasonable approximation to account for rotational
enhancement by simply adding a rotational band on each band head;
double-counting of states would not occur.

Rotational states are consequently taken into account by adding a
rotational band on top of each intrinsic band-head for deformed
nuclei (here defined as nuclei with calculated quadrupole
deformation $|\varepsilon_2|\geq 0.05$). The rotational energy of
the different angular momentum states, $I$, in the rotational band
is given by
\begin{equation}
  E_{\rm rot}^i(I,K_i) = \frac{I(I+1)-K_i^2}{2\mathcal{J}_\bot(\varepsilon_2,\Delta^i_{\rm
  p},\Delta^i_{\rm n})},
  \label{RotEnergy}
\end{equation}
where $K_i$ is the angular momentum projection on the symmetry axis
of the intrinsic state $i$ upon which the rotational band is built.
The quadrupole deformation is denoted by $\varepsilon_2$ while
$\Delta_{\rm p}^i$ and $\Delta_{\rm n}^i$ denote the proton and
neutron pairing gaps of the intrinsic state $i$. The moment of
inertia around an axis perpendicular to the symmetry axis
$\mathcal{J}_\bot(\varepsilon_2,\Delta^i_{\rm
  p},\Delta^i_{\rm n})$ is approximated by
the rigid-body moment of inertia with deformation $\varepsilon_2$, modified by
the calculated pairing gaps for the considered state, as given in
Ref.~\cite{MomInert}. Given the angular momentum
projection $K$ and parity $\pi$ of the band-head the rotational band includes
the following levels
\begin{equation}
  I^\pi = \left\{
    \begin{array}{ll}
      K^\pi, (K+1)^\pi,(K+2)^\pi,... & {\rm if~} K\neq 0,\\
      0^+,2^+,4^+,... & {\rm if~} K=0^+,\\
      1^-,3^-,5^-,... & {\rm if~} K=0^-.
    \end{array}
\right.
\end{equation}
The Coriolis anti-pairing effect is neglected and no virtual
crossings of rotational bands are taken into account. Thus, the
pairing gap and moment of inertia are assumed to be unchanged from
the band-head pairing gap for all states in the rotational band.
This approximation is reasonable since mainly low-spin states play a
role in the present study.

The total energy of the calculated level is thus given by
\begin{equation}
E_i = E_{\rm mb}^i + E_{\rm rot}^i(I,K_i)
\label{TotalEnergy}
\end{equation}
where $E_{\rm mb}^i$ is the energy, Eq.~(\ref{Total}), of the
intrinsic many-body configuration $i$, and $E_{\rm rot}^i(I,K_i)$ is
the rotation energy, Eq.~(\ref{RotEnergy}), of the level with
angular momentum $I$ built from the intrinsic many-body
configuration with angular momentum projection $K_i$. These energies
are used to calculate the level density according to
Eq.~(\ref{levdendef}).

\begin{figure}
  \includegraphics[clip=true,width=0.95\columnwidth]{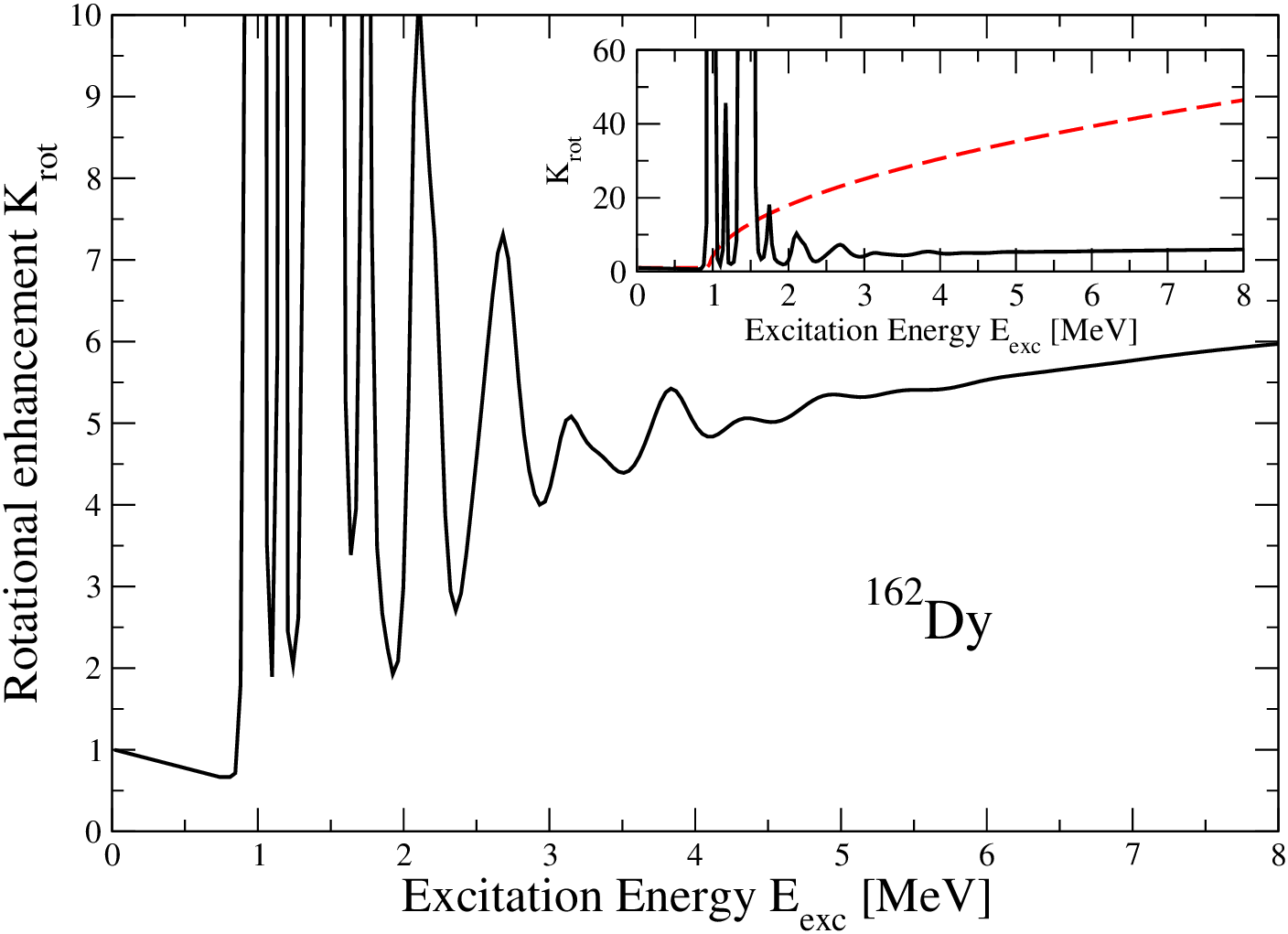}
  \caption{(Color online) Rotational enhancement, $K_{\rm rot}$, for the nucleus $^{162}$Dy
    as a function of excitation energy. The insert shows the enhancement
    compared to the simple enhancement model of Ref.~\cite{HansenJensen} (red
    dashed line).}
  \label{FIG:DyKrot}
\end{figure}

Fig.~\ref{FIG:DyKrot} shows the rotational enhancement, $K_{\rm
rot}$, for the well-deformed nucleus ($\varepsilon_2$=0.26)
$^{162}$Dy (cf.~Fig.~\ref{FIG:ALL3}) calculated as the ratio of the
level density when rotations are included or excluded. For low
excitation energies there are large fluctuations which are artifacts
of the low level density combined with the smoothing procedure of
Sec.~\ref{sec:spread}. For higher excitation energies ($\gtrsim 3$
MeV) the rotational enhancement is a slowly increasing function, of
the order of a factor 5 at the neutron separation energy. This
prediction is compared to the SU(3) model of
Ref.~\cite{HansenJensen}, which is shown in the insert of
Fig.~\ref{FIG:DyKrot}. The SU(3) model gives almost an order of
magnitude larger enhancement for excitation energies in the region
of the neutron separation energy. We note that a combinatorial
level-density model based on the Nilsson potential gives a
rotational enhancement for $^{162}$Dy similar to our results
\cite{Magne}.

%----------------------------------------------------

\subsection{Vibrations}
\label{sec:VIB}

Many nuclei exhibit low-lying states of vibrational character, which
are usually of quadrupole or octupole type. Such low-lying
vibrational states appear at excitation energies not much lower than
the 2qp excitations which in a coherent way build up the collective
state. We therefore believe that the vibrational enhancement of the
level density, contrary to the rotational enhancement discussed
above, must be described microscopically.

In order to describe vibrational states the
Quasi-particle-Tamm-Dancoff-Approximation (QTDA) is used. According
to the Brink-Axel hypothesis \cite{BrinkHyp,AxelHyp} phonons are
built on every intrinsic many-body configuration $E_{\rm mb}^i$. The
QTDA equation is solved for each state $i$ in order to get all
possible phonon excitation energies and wave-functions. This means
solving the QTDA equations millions of times for every single
nucleus.

The residual interaction is approximated by the double stretched
(isoscalar) Quadrupole-Quadrupole interaction. This interaction is
well defined in the case of a harmonic oscillator potential. In the
case of a finite-depth potential as the folded-Yukawa potential the
interaction should take into account additional finite size effects,
for example as is done in Ref.~\cite{KvasilVPM}. In the present work
the finite-depth effects are ignored and the double stretched
approach is used as defined in Refs.~\cite{Kishimoto,SvenQQ}.

The QTDA secular equation can be written \cite{Rowe}
\begin{equation}
  \frac{1}{\chi_{2K}} = \sum_{\mu \nu} \frac{\left|\left<\mu\left|
  \bar{Q}_{2K} \right|\nu\right>\right|^2 (U_\mu V_\nu+V_\mu
  U_\nu)^2 }{(E_\mu^{{\rm qp},i}+E_\nu^{{\rm qp},i})-(\hbar \omega)_j^i},
\label{QTDAEQ}
\end{equation}
where the effect of Eq.~(\ref{UVchange}) has not been explicitly written
out. $\bar{Q}_{2K}$ is the double stretched quadrupole operator, where the
components $K=0$ and $K=2$ are
considered. The set of roots $\{(\hbar \omega)_j^i\}$ of this equation is the excitation
energies of the vibrational phonons $j$ on top of the intrinsic state $i$, whereas the poles $E_\mu^{\rm
  qp,i}=\sqrt{(e_\mu-\lambda)^2+\Delta_i^2}$ are the unperturbed
two-quasiparticle excitations on the many-body configuration $E_{\rm
mb}^i$ with pairing gap $\Delta_i$ as calculated by
Eqs.~(\ref{BCSEQ1}) and (\ref{BCSEQ2}), and $e_\mu$ are the
single-particle energies.

The self-consistent coupling strength is given by \cite{Kishimoto}
\begin{equation}
  \chi_{2K} = \frac{8\pi }{5} \frac{M\omega_0^2}{A\left<
  \bar{r}^2\right>+g_{2K}\sqrt{\frac{4\pi}{5}}A\left< \bar{Q}_{20}\right>},
\label{coupstr}
\end{equation}
where $g_{20}=1$ and $g_{22}=-1$. The expectation-values $\left<
\bar{r}^2\right>$ and $\left<\bar{Q}_{20} \right>$ are calculated in
double stretched coordinates \cite{Kishimoto}. To test that this
gives overall reasonable result, we have verified for a large number
of nuclei that for the $K=0$ and the $K=2$ components of the
iso-scalar Giant Quadrupole Resonances the calculations agree well
with the systematics of the Giant Resonance energies $\hbar
\omega_{\rm GQR}=58A^{-1/3}$ MeV \cite{Kishimoto}.

The phonons are never repeated and double counting is explicitly
avoided by the following procedure. The phonon
wave functions are given by
\begin{equation}
  \mathcal{O}^\dagger = \sum_{\mu,\nu} X_{\mu,\nu} a_\mu^\dagger a_\nu,
\end{equation}
where $X_{\mu,\nu}$ are the wave-function components of all excited
quasi-particle states
$a_\mu^\dagger a_\nu$ on top of the configuration
$E_{\rm mb}^i$. The level density is increased by one state at the energy of the
phonon $(\hbar \omega)_j^i$, and decreased by the amount given by the
wave-function component $X^2_{\mu,\nu}$ at the energy of the corresponding
pole ($E_\mu^{{\rm qp},i}+E_\nu^{{\rm qp},i}$). The change in level density due to the phonon $j$ on top of intrinsic state $i$ is thus
\begin{multline}
  \delta \rho(E) = \delta (E-E_{\rm mb}^i-(\hbar \omega)_j^i)-  \\ \sum_{\mu,\nu} X_{\mu,\nu}^2
  \delta \left(E-E_{\rm mb}^i-(E^{{\rm qp},i}_\mu+E^{{\rm qp},i}_\nu ) \right),
  \label{levdenschange}
\end{multline}
where $E$ is the excitation energy relative to the ground-state.
This procedure increases the level density by reducing the strength
of the pure quasi-particle excitations and adding strength by means
of the phonons (which are pushed down in energy due to the
$QQ$-interaction).

Most phonons have little collectivity in the QTDA approximation, as
most of the phonon excitations lie very close to a pure
quasi-particle excitation, and hence the wave-function is completely
dominated by that single component. Even if there are a few very
collective low-lying phonon states the vibrational enhancement of
the level density becomes small, since most excited states provide
quite non-collective phonon excitations. Due to this
non-collectivity of most phonons there is no way that the phonons,
in general, can be repeated to form two- or even three-phonon
excitations.

\begin{figure}
  \includegraphics[clip=true,width=0.95\columnwidth]{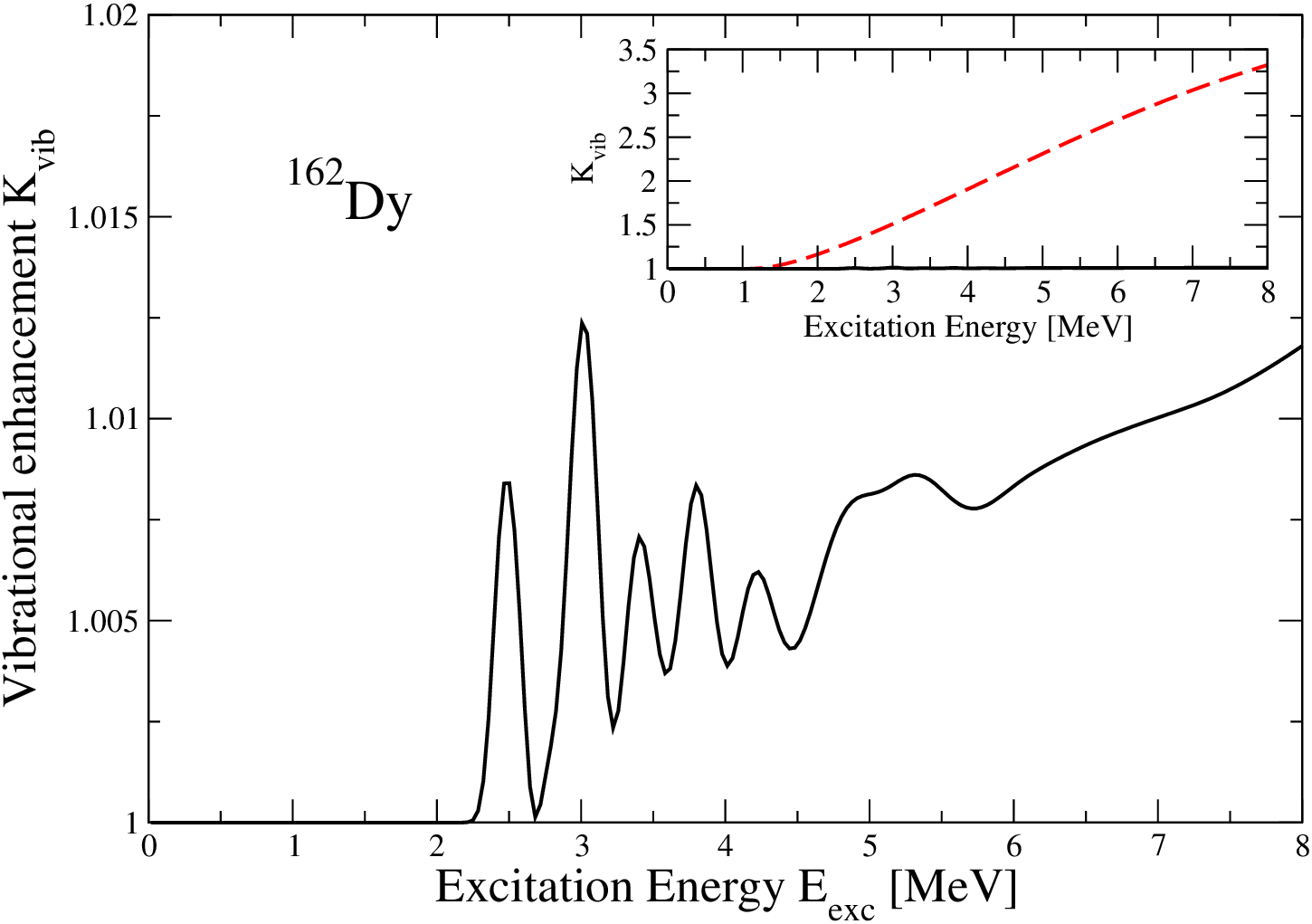}
  \caption{(Color online) Vibrational enhancement, $K_{\rm vib}$, using the QTDA method for the nucleus $^{162}$Dy
    as a function of excitation energy. The insert shows the enhancement
    compared to the enhancement of the attenuated phonon model of
    Ref.~\cite{GSM} (red dashed line).}
  \label{FIG:DyKvib}
\end{figure}

The vibrational enhancement factor in this method is in general
quite small, of the order of a few percent. This is substantially
lower than predictions from other models. For example, the
attenuated phonon method \cite{Goriely2006,GSM,IgnatyukBook} and the
Boson partition function method \cite{GorielyOSLO2008} both give up
to an order of magnitude enhancement at the neutron separation
energy. Fig.~\ref{FIG:DyKvib} shows the vibrational enhancement as a
function of excitation energy for $^{162}$Dy. The effect is very
small, close to 1 \% at 7 MeV excitation energy. For the same
nucleus the attenuated phonon method gives an enhancement factor of
about 3 as shown in the insert of Fig.~\ref{FIG:DyKvib}.

The level-density enhancement due to quadrupole vibrations is thus
found to be very small. Higher multipole vibrations (as octupole
vibrations) are expected to contribute with enhancements of the same
order of magnitude or less and are neglected in the CFY model.

%----------------------------------------------------

\subsection{Many-body damping width}
\label{sec:spread}
In the mean-field approach all excited many-body states are treated as
non-interacting. A
residual two-body interaction will mix the many-body states obtained from the
combinatorics. Smearing effects from the residual interaction can
approximately be taken into account by
assuming a spreading width of all excited states. The spreading width is
implemented in terms of a Gaussian envelope with width $\sigma$, i.e. the delta
functions in Eq.~\ref{levdendef} are replaced by Gaussians. Estimates of the
spreading width FWHM gives \cite{Lauritzen}
\begin{equation}
  \Gamma = 0.039 \left(\frac{A}{160} \right)^{-1/2} E^{3/2}~{\rm MeV},
\end{equation}
where the FWHM is related to the Gaussian spreading width
$\sigma=\frac{\Gamma}{2\sqrt{2\ln 2}}$.

Assuming that all excited many-body states in an energy bin $\Delta
E$ are uniformly distributed, the level density becomes
\begin{multline}
  \rho(E_b) = \sum_a \rho(E_a) \frac{1}{2} \left[{\rm erf} \left(\frac{E_a+\Delta
  E/2-E_b}{\sqrt{2}\sigma} \right) \right. -\\ -\left. {\rm erf} \left(\frac{E_a-\Delta
  E/2-E_b}{\sqrt{2}\sigma} \right) \right],
\end{multline}
for bin-point $b$. The method implies a smearing out of
level-density properties over a range  $\Gamma$, which is smoothly
increasing with excitation energy. As a consequence fluctuations of
energies and wave-functions in the range $\Gamma$ follow GOE
statistics of Random Matrix theory, that is often denoted as quantum
chaos in the nucleus, see e.g Refs.~\cite{Aberg90,Zelevinsky96}.

%%%%%%%%%%%%%%%%%%%%%%%%%%%%%%%%%%%%%%%%%%%%%%%%%%%%%%%%%%%%%%%

\section{Level Distributions}
\label{sec:leveldistr}

In the present combinatorial approach it is possible to extract
different distributions exhibiting details of the level-density
function. We present here microscopically calculated distributions
for pairing gaps (Sec.~\ref{sec:Pairgap}), parity (Sec.~\ref{Parity
distr}), and angular momentum (Sec.~\ref{sec:spindistr}).

%-----------------------------------------------

\subsection{Pairing-gap distribution}
\label{sec:Pairgap} 
\begin{figure}
  \includegraphics[clip=true,width=0.95\columnwidth]{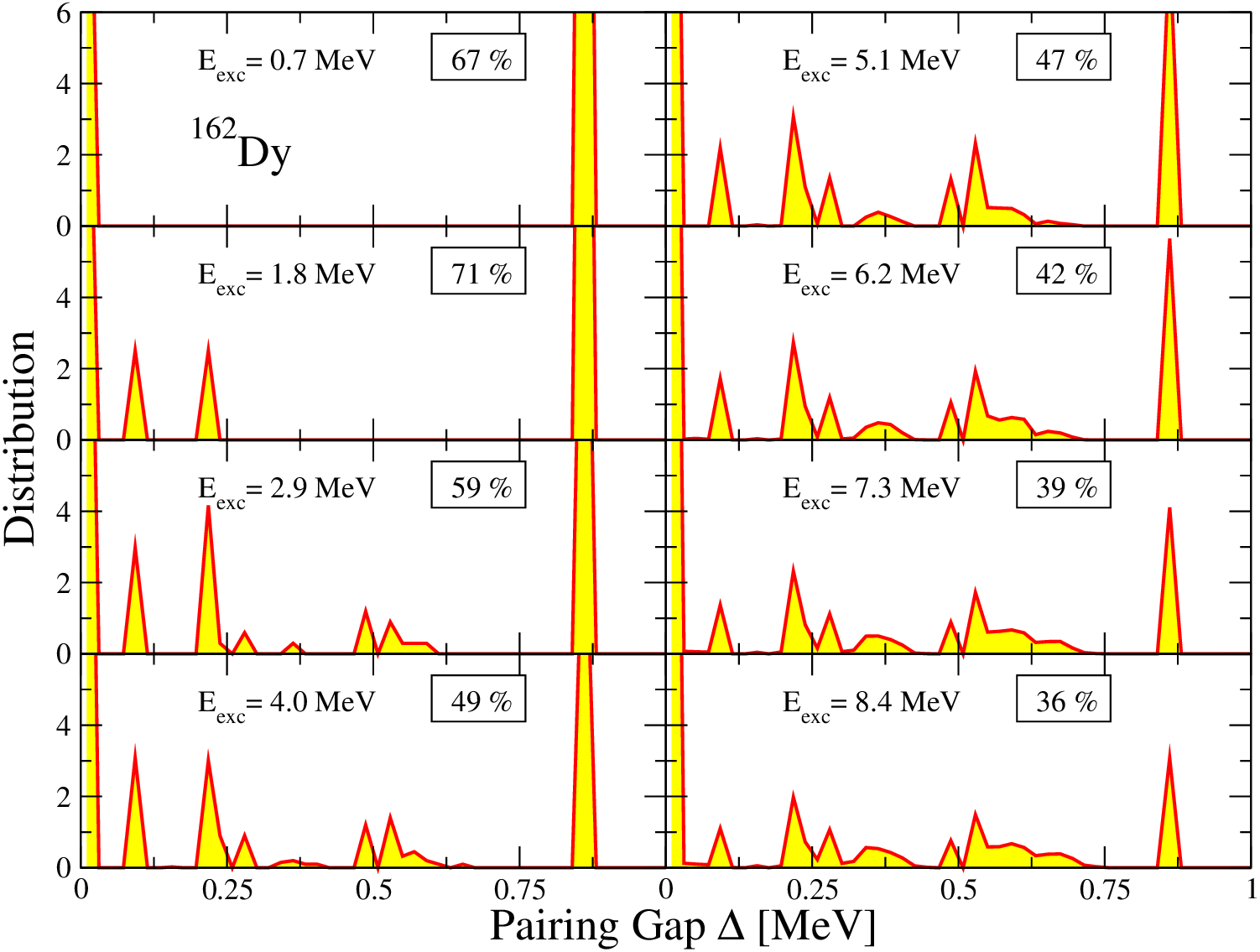}
  \caption{(Color online) Proton pairing-gap distributions at different excitation energies
    $E_{\rm exc}$ for $^{162}$Dy shown in 20 keV pairing-gap bins. The proportion of
    paired states ($\Delta>0$) are shown in percent in the boxes. }
  \label{FIG:Pairdistr}
\end{figure}

\begin{figure}
  \includegraphics[clip=true,width=0.95\columnwidth]{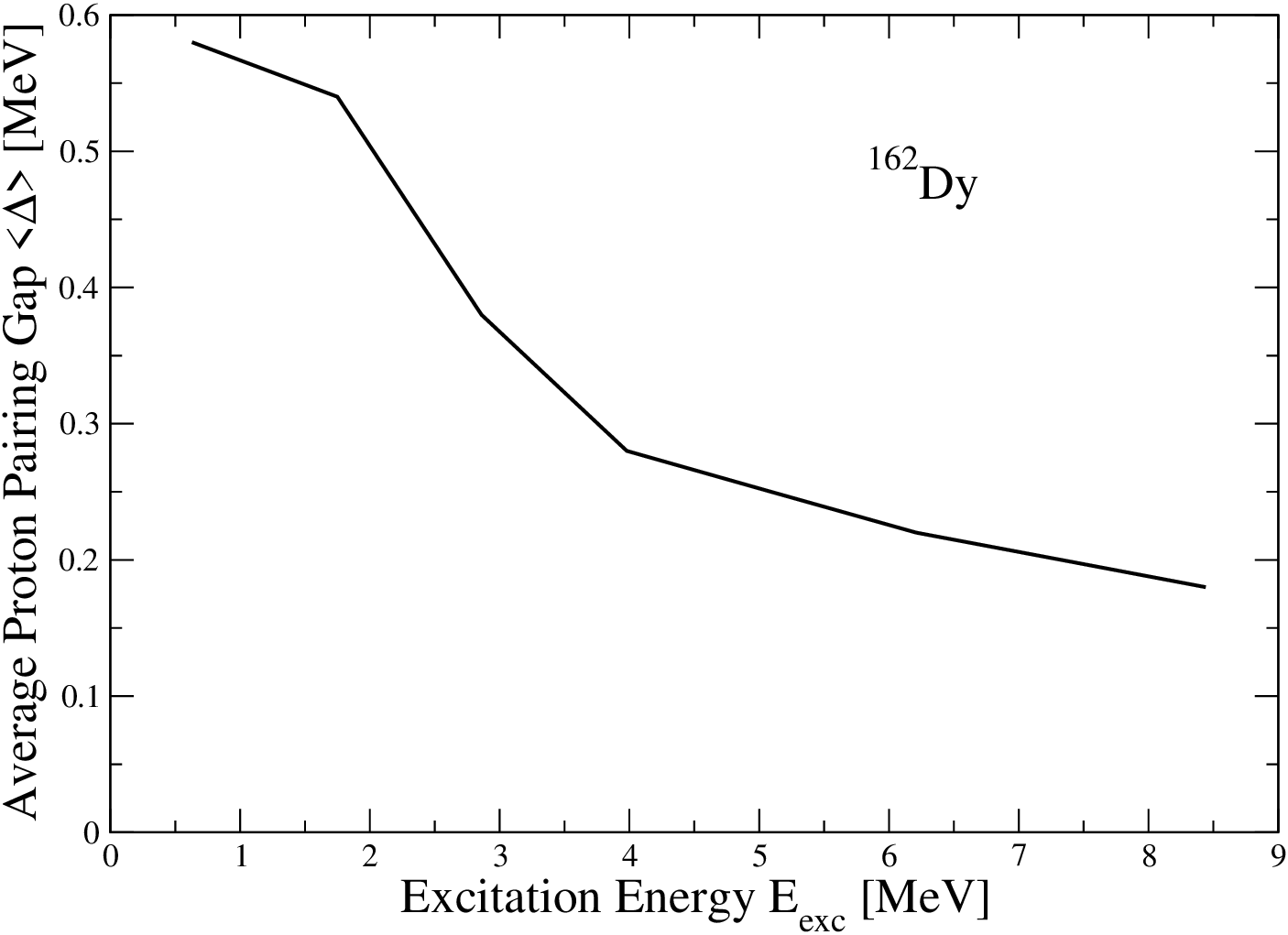}
  \caption{Average proton pairing gap $\left< \Delta\right>$ at different
    excitation energies $E_{\rm exc}$ for $^{162}$Dy. The mean values are
    given by the
    distributions in Fig.~\ref{FIG:Pairdistr}.}
  \label{FIG:PairMean}
\end{figure}

The BCS equations, Eqs.~(\ref{BCSEQ1}) and
(\ref{BCSEQ2}), are solved for all individual many-body
configurations and hence pairing gaps for all states are obtained.
In Fig.~\ref{FIG:Pairdistr} the distribution of proton pairing gaps
for $^{162}$Dy is shown for a number of excitation energies. For the
lowest excitation energies only the ground-state and the states in
the ground-state rotational band exist. The pairing gap of the
levels in the rotational band is fixed to be the same as the pairing
gap of the band-head, see Sec.~\ref{sec:ROT}. As the excitation
energy increases levels with reduced pairing gaps appear. However,
no transition to a completely unpaired system is observed, and
levels with non-collapsed gaps ($\Delta>0$) survive to the highest
considered excitation energies. At the highest considered excitation
energy for $^{162}$Dy, $E_{\rm exc}=8.4$~MeV, $36\%$ of the levels still
have a non-zero proton pairing gap, see Fig.~\ref{FIG:Pairdistr}.
The mean value of the proton pairing gap $\left< \Delta \right>$ at
different excitation energies is shown in Fig.~\ref{FIG:PairMean}
for $^{162}$Dy. Between 2 MeV and 3.5 MeV excitation energy there is
a rapid decrease in the mean pairing gap. At higher excitation
energies the decrease is slower. At 8.4 MeV excitation energy the
mean value is $\left< \Delta \right>=0.2$~MeV. The reason why
pairing may survive and be of substantial size in states at these
high energies can be explained as follows: Several highly excited
states are built by exciting particles and holes far from the Fermi
energy, where blocking in the BCS equations has no effect on the
pairing gap.

The non-collapsed pairing gaps influence the moment of inertia and
keep it reduced as compared to the rigid-body value. This implies
that even at excitation energies in the region of the neutron
separation energy the moment of inertia is on average smaller than
the rigid-body value. A similar observation is made in
Ref.~\cite{HilaireGirod}.

%----------------------------------------------------

\subsection{Parity distribution}
\label{Parity distr} In  Fermi-gas level-density models there is an
implicit assumption of equal number of states with different parity
at any given excitation energy. However, models that take into
account microscopic effects show  clear structure in the parity
ratio, see eg. Refs.~\cite{AlhassidParity,GrimesParity,Cerf}. In
connection with astrophysical reaction rates the parity ratio may
play an important role, and can be included in the Hauser Feshbach
formalism, see eg Refs.~\cite{Loens,Mocelj}.

\begin{figure}
  \includegraphics[clip=true,width=0.95\columnwidth]{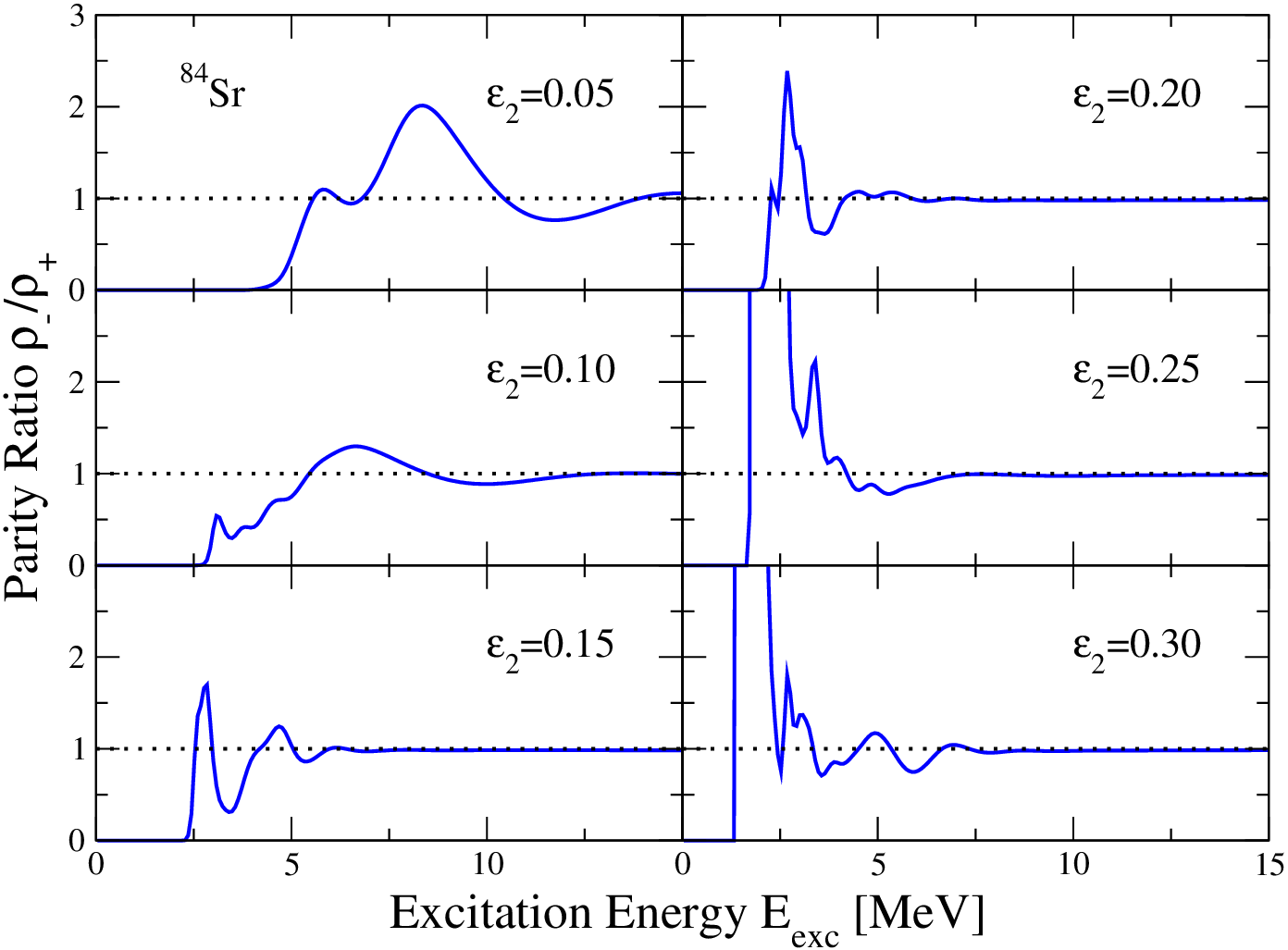}
  \caption{(Color online) Parity ratio versus excitation energy for $^{84}$Sr calculated with
    different deformations. Large deformations imply a
    parity distribution close to 1 even at low excitation energies. For small
    deformations the parity non-equilibrium can survive to high excitation
    energies. }
  \label{FIG:DefDep}
\end{figure}

From the single-particle point of view it is clear that the parity
ratio should show structure. For $^{84}_{38}$Sr$_{46}$ the situation
is illuminating. The nucleus is close to spherical in its
ground-state ($\varepsilon_2=0.05$) \cite{MollerMass}. It has 2
proton holes in the $pf$ shell and 6 neutrons in the $g_{9/2}$
shell. Since there are large gaps in the single-particle spectrum
the energy to excite nucleons across the gaps, which cause changes
in parity, is quite large. The parity ratio displays long-range
oscillations, see upper left panel of Fig.~\ref{FIG:DefDep}, which
are directly connected to the large single-particle gaps. The
single-particle gaps effectively decrease with increasing
deformation, and the oscillations are therefore expected to decrease
with deformation. To test these ideas the parity ratio in $^{84}$Sr
is calculated at different deformations and shown in
Fig.~\ref{FIG:DefDep}. Indeed, for larger deformations the
oscillatory pattern vanishes and the parity ratio equilibrates to
one.

\begin{figure}
  \includegraphics[clip=true,width=0.95\columnwidth]{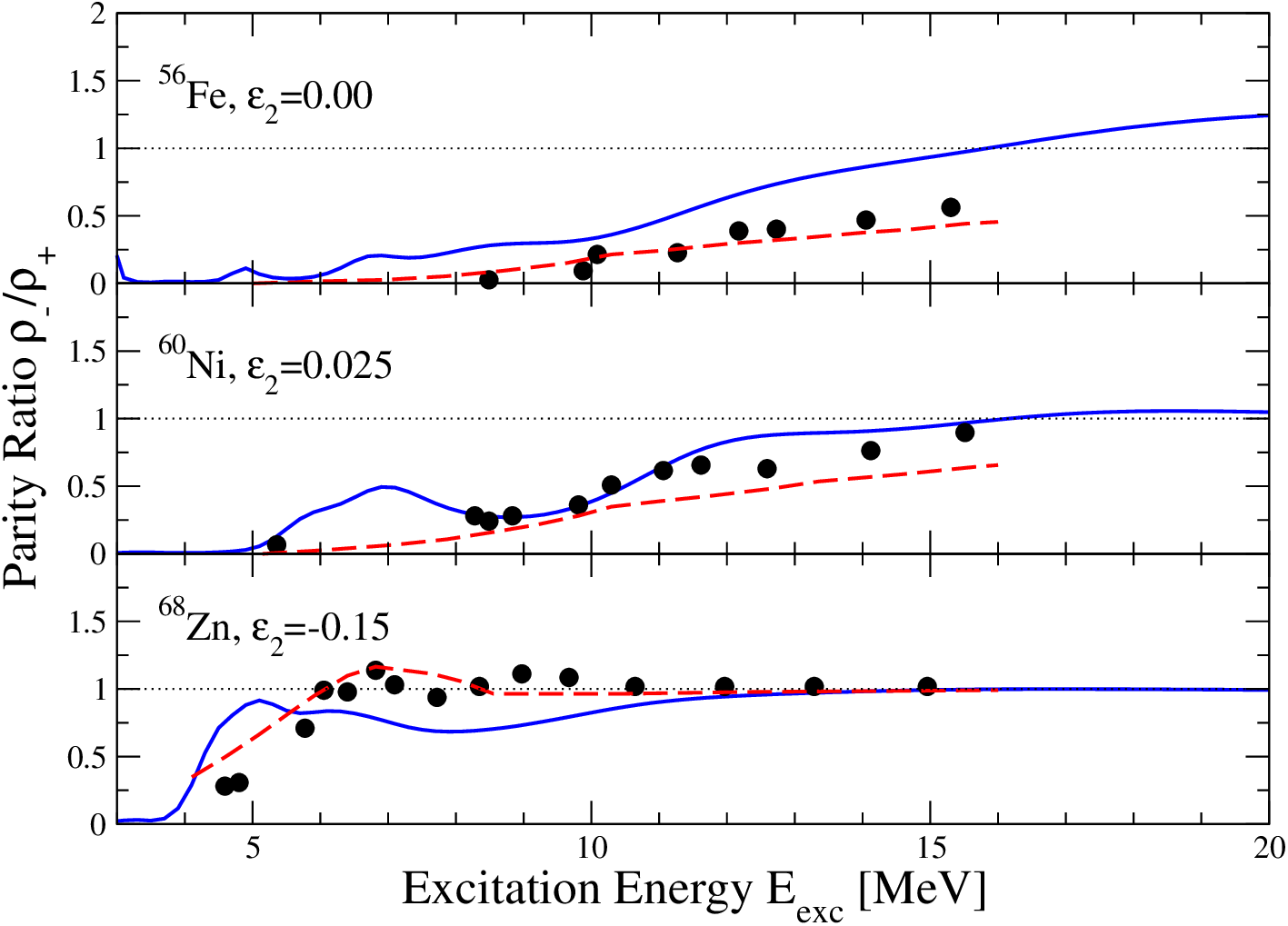}
  \caption{(Color online) Parity ratio versus excitation energy. The solid blue lines are
    calculated with the CFY model. The red dashed line and
    black dots are given by a statistical model and a Monte-Carlo method,
    respectively \cite{AlhassidParity}. }
  \label{FIG:FeNiZn}
\end{figure}

In Fig.~\ref{FIG:FeNiZn} the parity ratios for $^{56}$Fe, $^{60}$Ni
and $^{68}$Zn are shown and compared to two calculations of
Ref.~\cite{AlhassidParity}. The solid blue lines show the CFY model,
the red dashed lines show the statistical parity projection model
and the black dots show calculations using the Shell-Model
Monte-Carlo method. The parity ratio in the Fe and Ni isotopes is
not equilibrated below $15$~MeV in any of the calculations, while
$^{68}$Zn equilibrates at much lower energies (below $10$ MeV). For
$^{60}$Ni and $^{68}$Zn there is a clear oscillatory behavior prior
to equilibration in the CFY model, and in the case of Ni there is a
good agreement between the Monte-Carlo method and the CFY model,
especially in the region of 8--12 MeV excitation energy. Note,
however, that fluctuations seen in the present micro-canonical
approach may be smeared out by the grand-canonical approach as used
in Ref.~\cite{AlhassidParity}.

\begin{figure}
  \includegraphics[clip=true,width=0.95\columnwidth]{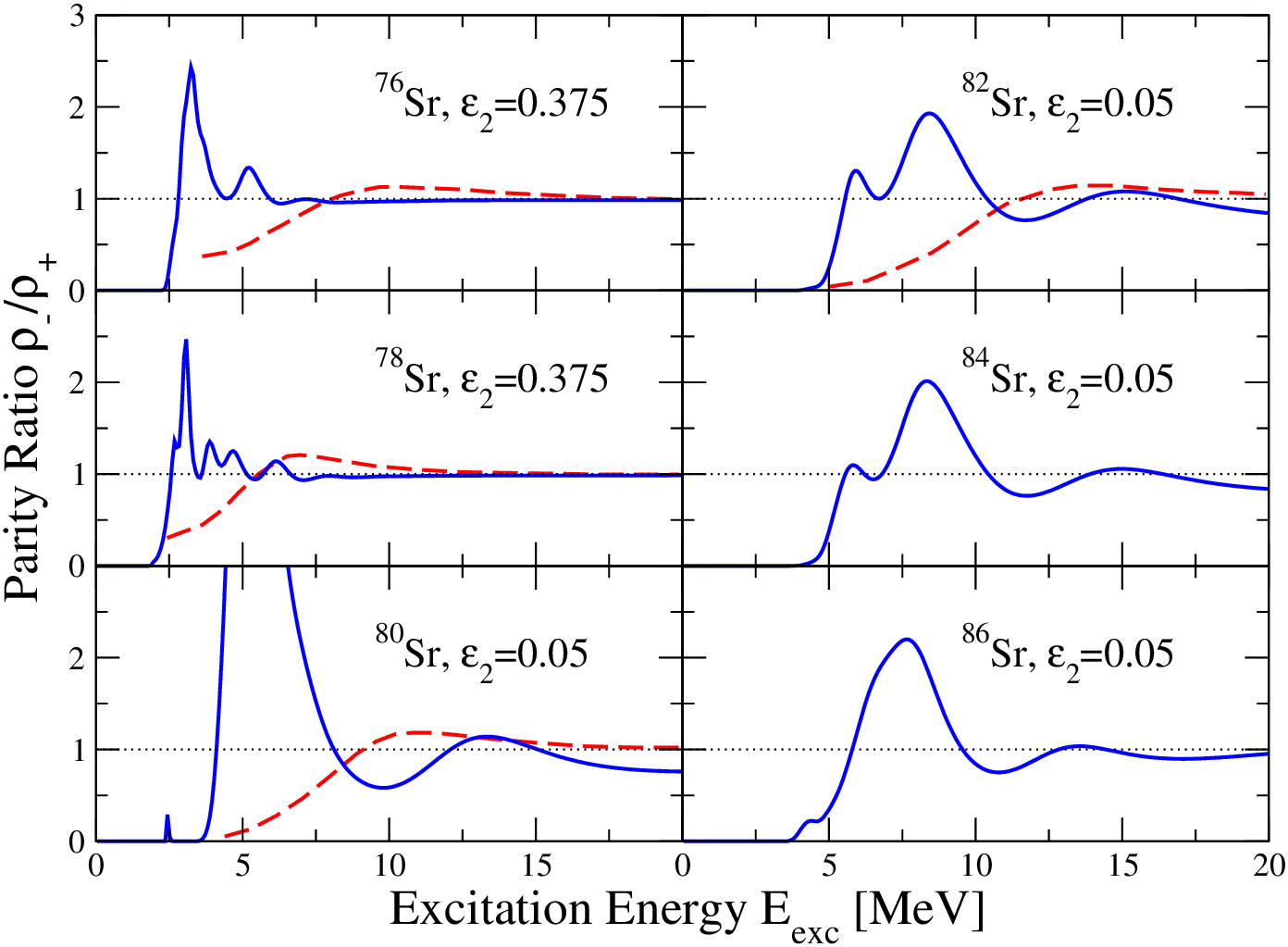}
  \caption{(Color online) Parity ratio versus excitation energy for Sr isotopes. The solid
    blue lines are obtained from the CFY model and the dashed red lines are
    obtained from the statistical
    parity projection of Ref.~\cite{Mocelj}.}
  \label{FIG:Sr}
\end{figure}

The parity ratio has been calculated for Sr-isotopes within a
statistical method in Ref.~\cite{Mocelj} where the impact on
astrophysical reaction rates was investigated and found to be small.
The statistical method gives at most one oscillatory maximum before
it equilibrates, as seen in Fig.~\ref{FIG:Sr} (dashed lines). In the
CFY model the parity ratio has substantially more structure (solid
lines). Nuclei with small ground-state deformations show long-range
oscillations which survive to high excitation energies before
equilibration. The overall results are quite different from the
model of Ref.~\cite{Mocelj}.

%----------------------------------------------------

\subsection{Angular momentum distribution}
\label{sec:spindistr}
In Fermi-gas models the distribution of angular momentum is given by
the Gaussian envelope in the spin cutoff model, see eg. Ref.~\cite{Rauscher},
\begin{equation}
  \mathcal{F}(U,I) = \frac{2I+1}{2\sigma^2} \exp
  \left(\frac{-I(I+1)}{2\sigma^2} \right),
  \label{GaussAngMom}
\end{equation}
which is obtained by random coupling of uncorrelated spins of the
nucleons. In Eq.(\ref{GaussAngMom}) the spin cutoff factor,
$\sigma$, is defined by
\begin{equation}
  \sigma^2 = \frac{\mathcal{J}_{\rm rigid}}{\hbar^2} \sqrt{\frac{U}{a}},
\end{equation}
where $\mathcal{J}_{\rm rigid}$ is the rigid-body moment of inertia,
$U=E-\delta$ is the effective excitation energy shifted by the back-shift
$\delta$, and $a$ is the level-density parameter.

\begin{figure}
\centering
\subfigure
{
  \includegraphics[clip=true,width=0.95\columnwidth]{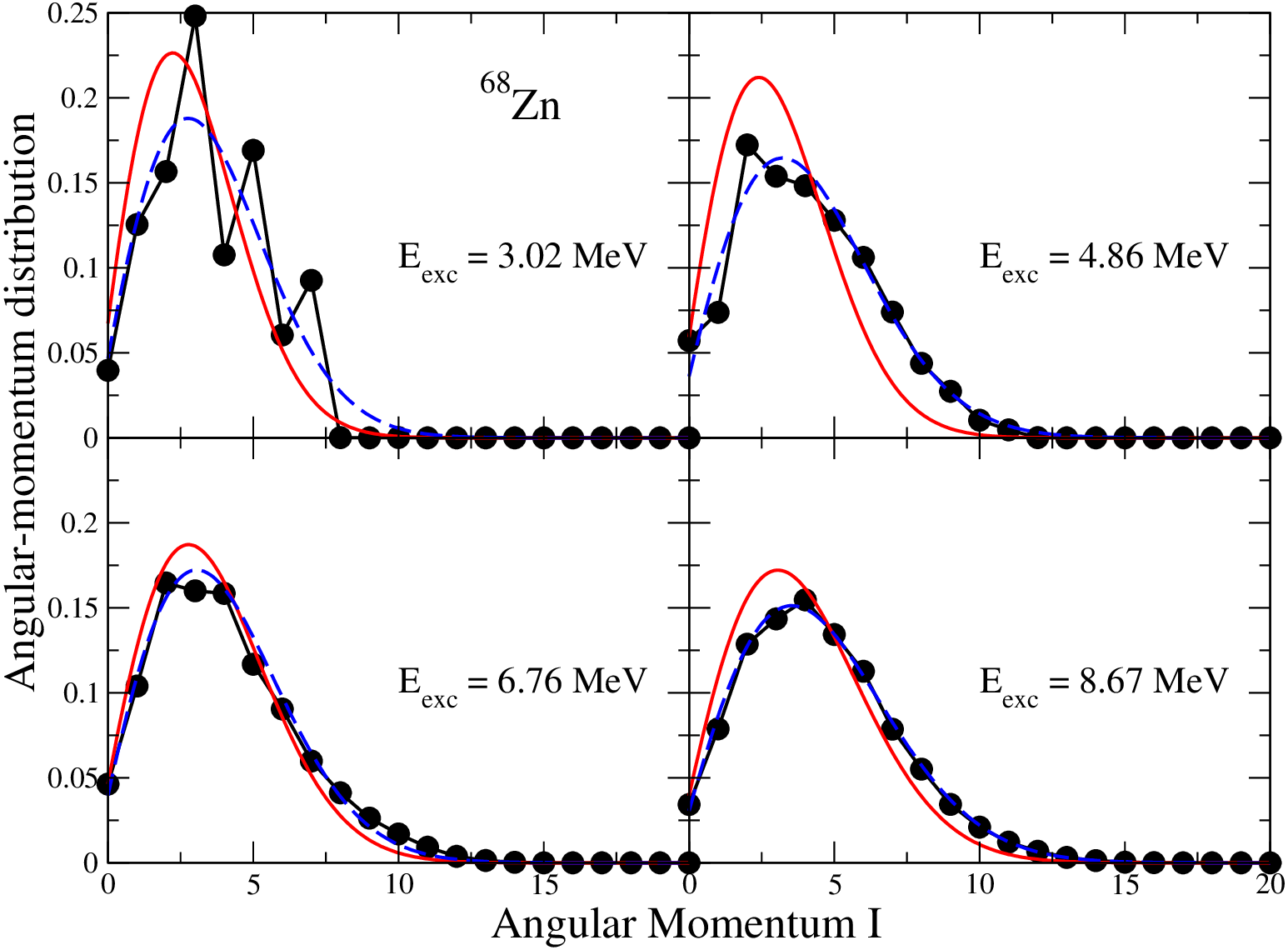}
%  \label{FIG:ZnSpinDist}
}
\subfigure
{
  \includegraphics[clip=true,width=0.95\columnwidth]{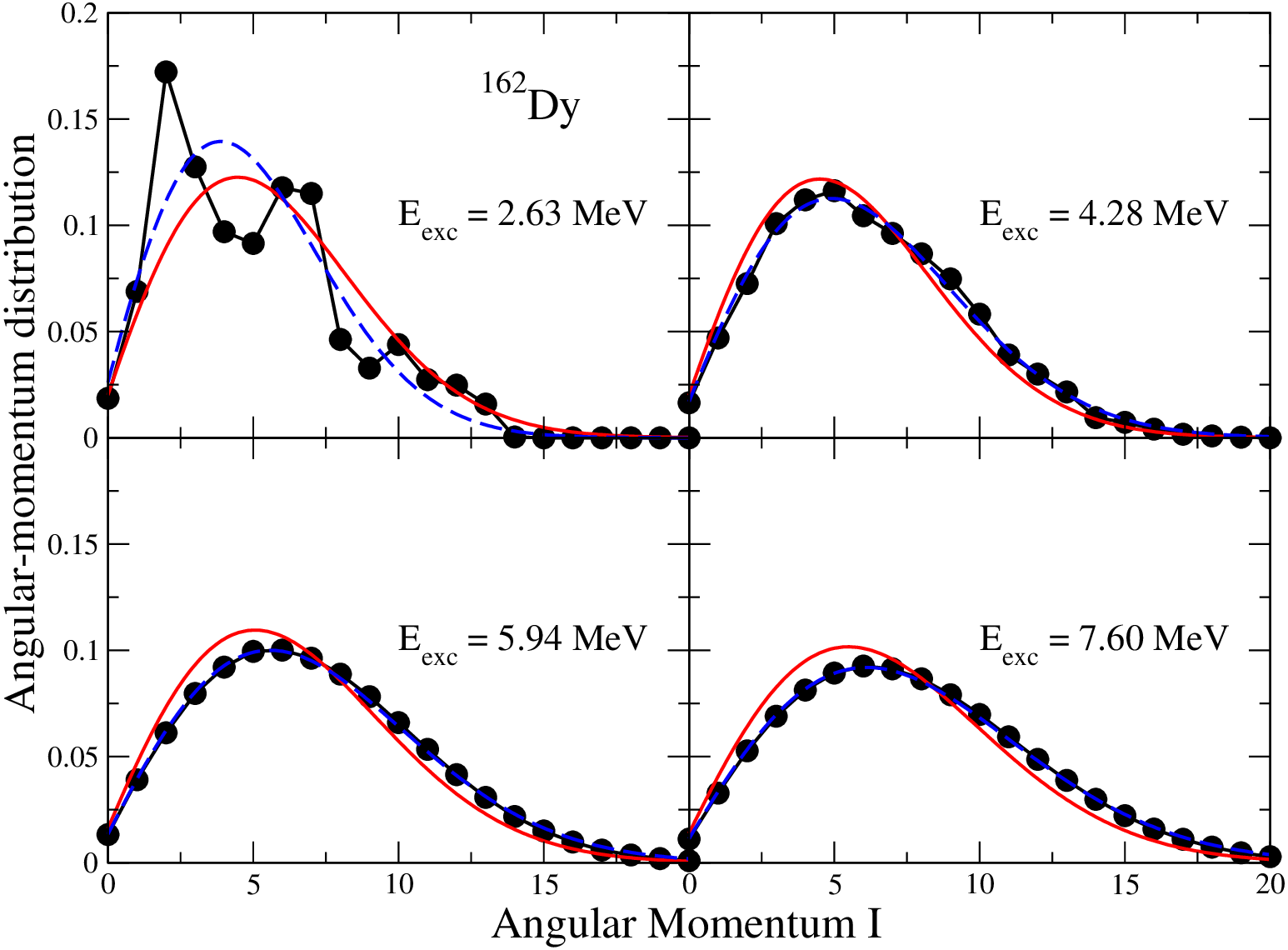}
%  \label{FIG:DySpinDist}
}
\caption{(Color online) Angular-momentum distribution for $^{68}$Zn and $^{162}$Dy in 4 different
  excitation-energy regions. The black lines with dots show results from the CFY
  model. The red solid lines and blue dashed lines show the Gaussians given
  by Eq.~(\ref{GaussAngMom}) using the spin cutoff factors from
  Ref.~\cite{Rauscher} and a direct fit to CFY, respectively. }
\label{FIG:SpinDist}
\end{figure}

In Fig.~\ref{FIG:SpinDist} the angular-momentum distributions for
$^{68}$Zn and $^{162}$Dy are shown for several excitation-energy
regions. The black lines with dots show the CFY model results while
the red solid lines show the Gaussian distribution of
Eq.~(\ref{GaussAngMom}) with the spin cutoff factors given in
Ref.~\cite{Rauscher}, and the blue dashed lines show Gaussian
distributions fitted to the combinatorial calculation. For low
excitation energies there are clear deviations from the Gaussian
profiles while for higher excitations the combinatorial distribution
tends to the Gaussian profile. However, for $^{68}$Zn the deviations
from a Gaussian angular momentum distribution remain up to about the
neutron separation energy ($S_n$=6.48 MeV).

For low excitation energies in $^{68}$Zn there is an odd-even spin
staggering which is not explained by the spin cutoff model. This
effect has also been observed in Fe-isotopes in the Shell Model
Monte-Carlo Method \cite{AlhassidSpinDist}.

\begin{figure}
  \includegraphics[clip=true,width=0.95\columnwidth]{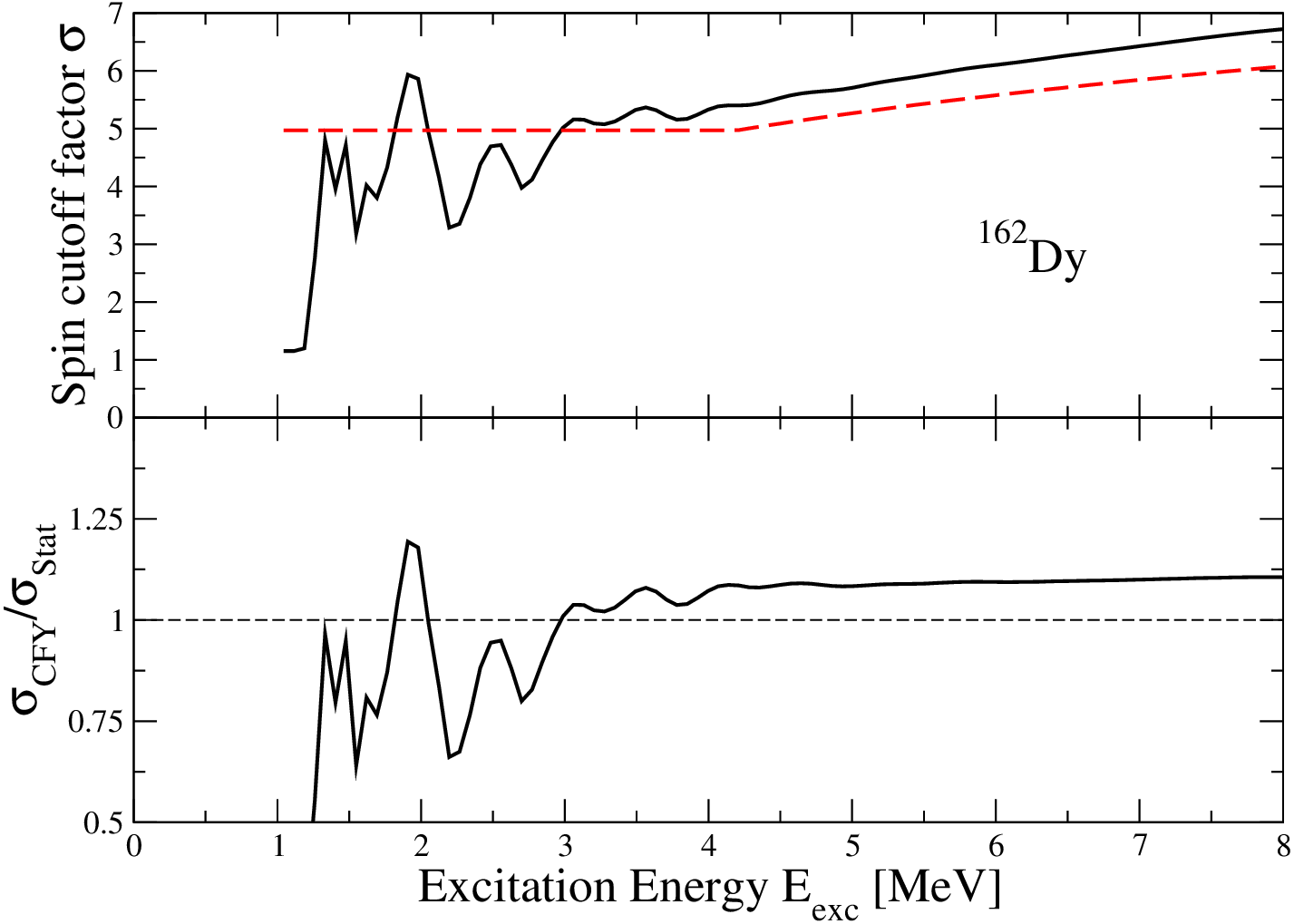}
  \caption{(Color online) The top panel shows the spin cutoff factor as a function of excitation
    energy for $^{162}$Dy in the CFY model (black solid line) and the
    statistical model of Ref.~\cite{Rauscher} (red dashed line). The bottom
    panel shows the ratio of the combinatorial and statistical spin cutoff
    factors of the top panel.}
  \label{FIG:Sigfit}
\end{figure}

Fig.~\ref{FIG:Sigfit} shows the spin cutoff factor deduced from the
CFY model (black solid line) and from the statistical model
\cite{Rauscher} (red dashed line), as a function of excitation
energy for $^{162}$Dy. For this particular nucleus the spin cutoff
factor for excitation energies $\gtrsim 3$ MeV is similar in shape
but \mbox{$\sim$10~\%} larger than in the statistical model
\cite{Rauscher}. The fact that the statistical model is smaller than
the CFY model, despite the presence of pairing in the CFY model, is
accidental for this nucleus. Since both the rigid body moment of
inertia and the level density parameter are fitted to experiments
using only three parameter, the detailed description for a
particular nucleus might give this result when compared to a very
different model as the CFY model.

%%%%%%%%%%%%%%%%%%%%%%%%%%%%%%%%%%%%%%%%%%%%%%%%%%%%%%%%%%%%%%%

\section{Comparison with Experimental data}
\label{sec:EXP}

%----------------------------------------------------

\subsection{Neutron separation-energy level spacings}
\label{sec:neutres}
The s-wave neutron resonance spacings constitute the most comprehensive
experimental database for comparison with NLD calculations
\cite{RIPL2}. This database serves as a benchmark for all large-scale
level-density models \cite{Goriely2008,Goriely2006,Rauscher,HilaireGirod}.

\begin{figure}
  \includegraphics[clip=true,width=0.95\columnwidth]{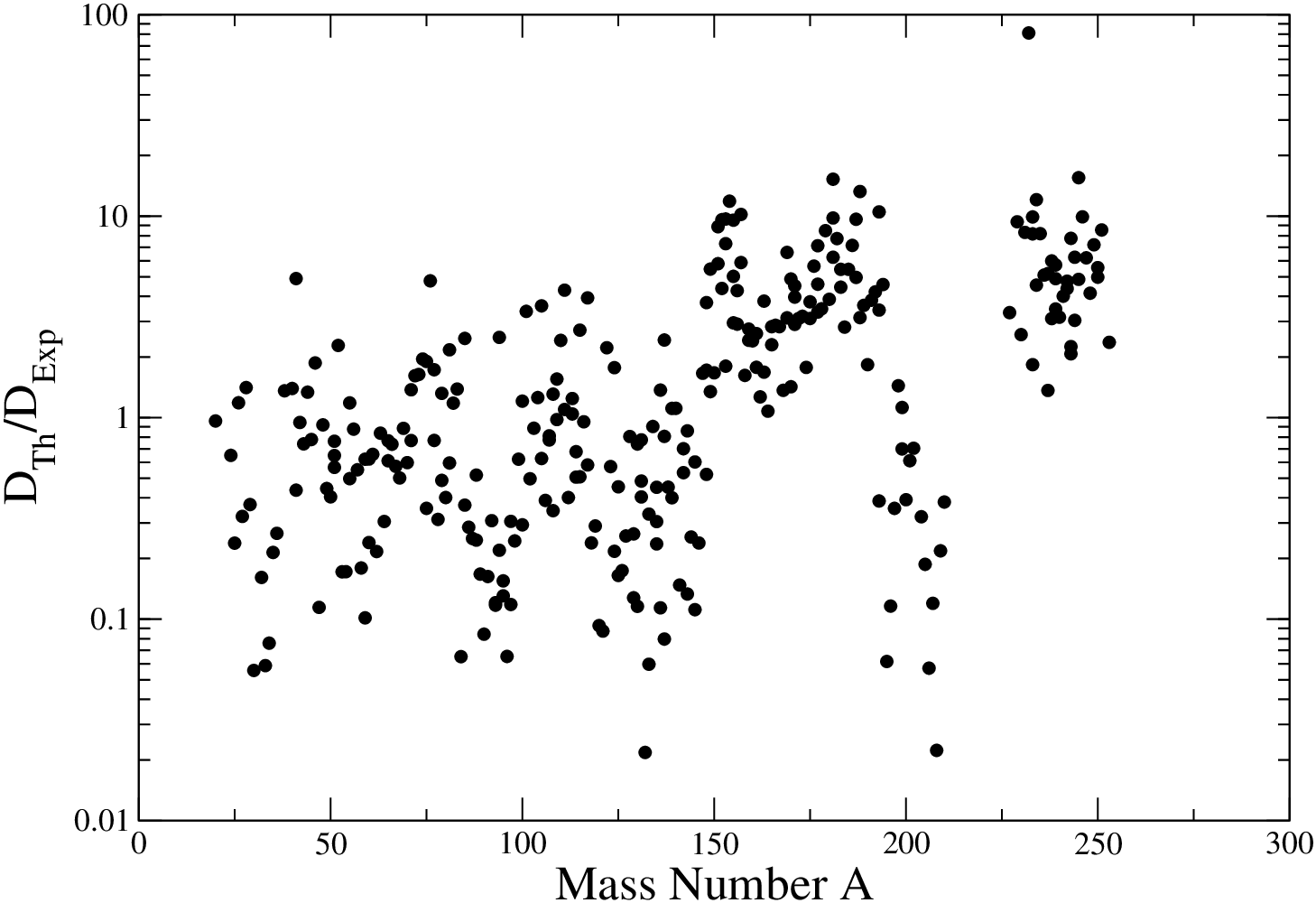}
  \caption{Ratio between theoretical and experimental level spacings at
   neutron separation energy
   versus mass number $A$. The rms-factor is $f_{\rm rms}=4.2$.
  The experimental data are taken from Ref.~\cite{RIPL2}.}
  \label{FIG:NeutSep}
\end{figure}

The s-wave neutron resonance spacing $D_0$ at the neutron separation
energy $S_{\rm n}$ of the compound nucleus $(Z,N)$ is obtained from
calculated level densities $\rho(E,I,\pi)$ as,
\begin{multline}
  \frac{1}{D_0} =\\ \left\{
    \begin{array}{ll}
      \rho\left(S_{\rm n},I_0+1/2,\pi_0 \right)+\rho \left(
      S_{\rm n},I_0-1/2,\pi_0\right) & {\rm for~} I_0>0,\\
      \rho\left(S_{\rm n},1/2,\pi_0 \right) & {\rm for~} I_0=0,\\
    \end{array}
  \right.
\end{multline}
where $I_0$ is the ground-state spin and $\pi_0$  is the
ground-state parity of the target nucleus $(Z,N-1)$. In
Fig.~\ref{FIG:NeutSep} we study the ratio of calculated and
experimental level spacings at the neutron separation energy,
$D_{\rm Th}/D_{\rm Exp}$, for all nuclei in the database.

The quality of a level-density model can be estimated by the
rms-factor \cite{HilaireGirod}
\begin{equation}
  f_{\rm rms} = \exp \left[ \frac{1}{N_e} \sum_{i=1}^{N_e} \ln^2
  \frac{D_{\rm Th}^i}{D_{\rm Exp}^i} \right]^{1/2},
\end{equation}
and the mean factor
\begin{equation}
  m = \exp \left[ \frac{1}{N_e} \sum_{i=1}^{N_e} \ln
  \frac{D_{\rm Th}^i}{D_{\rm Exp}^i} \right],
\end{equation}
where $D_{\rm Th}^i$ and $D_{\rm Exp}^i$ are the theoretical and
experimental level spacings and $N_e$ is the number of nuclei in the
database. The CFY model gives $f_{\rm rms}=4.2$ and $m=1.1$, see
Fig.~\ref{FIG:NeutSep} of the error ratio versus $A$.  These results are compared to other statistical and
combinatorial models in Table~\ref{RMStable} and discussed in
Sec.~\ref{sec:othermodels}.

There seems to be a clear residual shell structure in the level
spacings, as seen in Fig.~\ref{FIG:NeutSep}, especially in the
doubly magic $^{208}$Pb region. A similar effect can be observed for
the Gogny model in Ref.~\cite{HilaireGirod}, and in results of the
Skyrme model with the BSk9 interaction, as shown in
Ref.~\cite{Goriely2006}. In addition, there seems to be a drift with
mass number where especially nuclei in the rare-earth and actinide
regions over-estimate the neutron separation level spacings. A
similar drift can be seen in the Gogny model of
Ref.~\cite{HilaireGirod} while it seems not to be present in the
Skyrme-HFB model in Ref.~\cite{Goriely2006}.

\begin{figure}
  \includegraphics[clip=true,width=0.95\columnwidth]{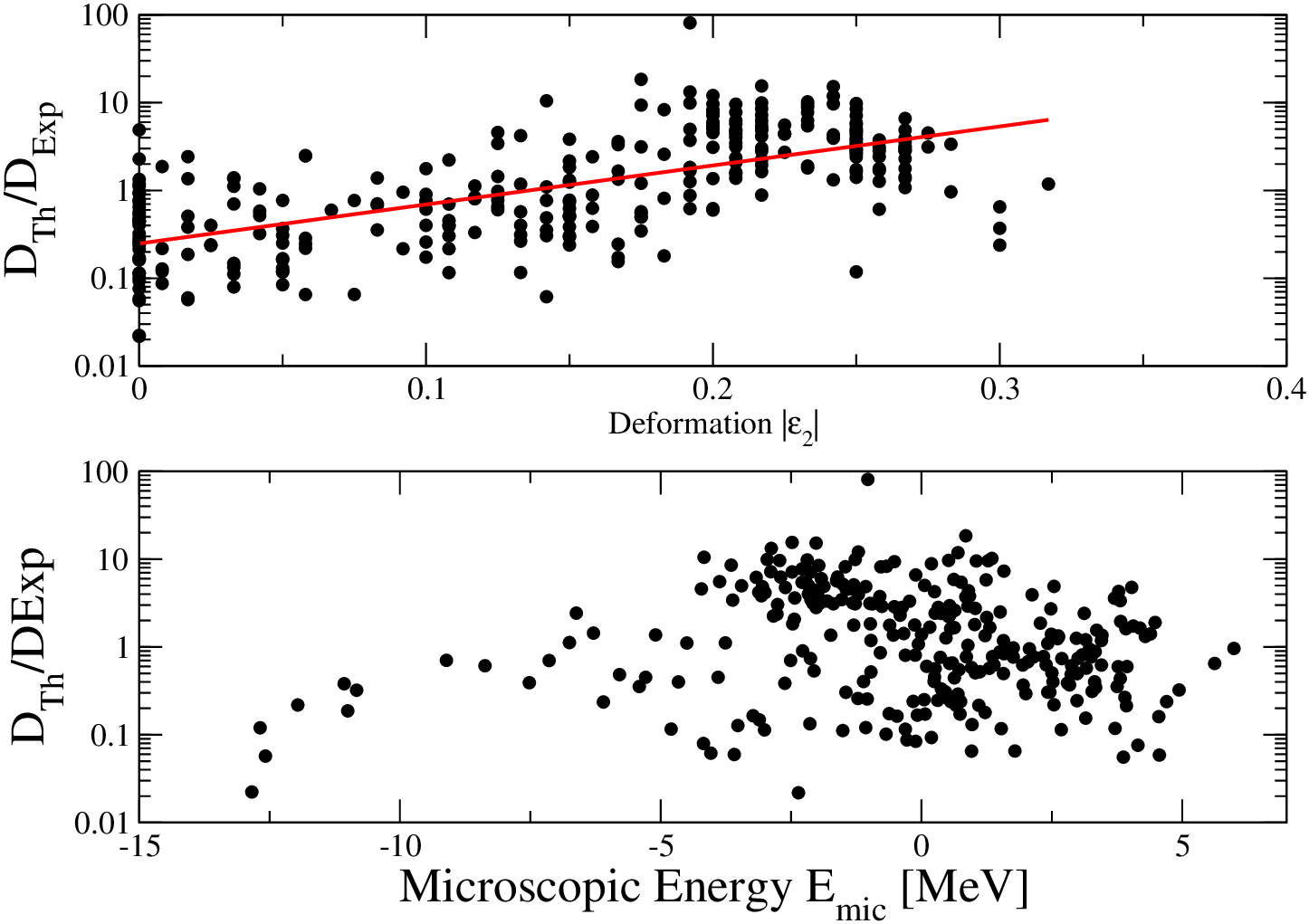}
  \caption{(Color online) The top panel shows
    the ratio  between theoretical and experimental level spacings at
    neutron separation energy versus the absolute value
    $|\varepsilon_2|$ of the
    calculated quadrupole deformation.
    The bottom panel shows this ratio as a function of the microscopic
    energy. The solid red line shows an exponential fit to illustrate the
    correlation. The microscopic energies and the quadrupole deformations
     are taken from
    Ref.~\cite{MollerMass}. The experimental data are from Ref.~\cite{RIPL2}.}
  \label{FIG:NeutSepDEF}
\end{figure}

The rare-earth and actinide regions, where too large level spacings
are calculated, mainly correspond to deformed nuclei. On the other
hand, nuclei corresponding to regions with particularly low ratio
$D_{\rm Th}/D_{\rm Exp}$ seen in Fig.~\ref{FIG:NeutSep} (around
$A$=32, 132 and 208) correspond to spherical nuclei. The
overestimation of $D$ for deformed nuclei (too low calculated level
density), and underestimation of $D$ for spherical nuclei (too high
calculated level density) seems to be approximatively systematic, as
seen in Fig.~\ref{FIG:NeutSepDEF}, where the ratio between
theoretical and experimental level spacings is shown versus the
absolute value of the deformation.

However, as seen in the lower panel of Fig.~\ref{FIG:NeutSepDEF}
there is no clear correlation with the microscopic energy $E_{\rm
mic}$ of Ref.~\cite{MollerMass} as might be expected from the
residual shell structure seen in Fig.~\ref{FIG:NeutSep}.

%----------------------------------------------------

\subsection{Detailed level-density functions}
The Oslo method is a commonly used experimental method for
extracting detailed level-density functions for large ranges of
excitation energy \cite{OsloMethod}. It provides a valuable test for
NLD models. However, the approach is  model dependent since the level
density is extracted by use of a back-shifted Fermi Gas approximation
and a spin-cutoff factor, see also discussion in
Ref.~\cite{GorielyOSLO2008}.

\begin{figure}
  \includegraphics[clip=true,width=0.95\columnwidth]{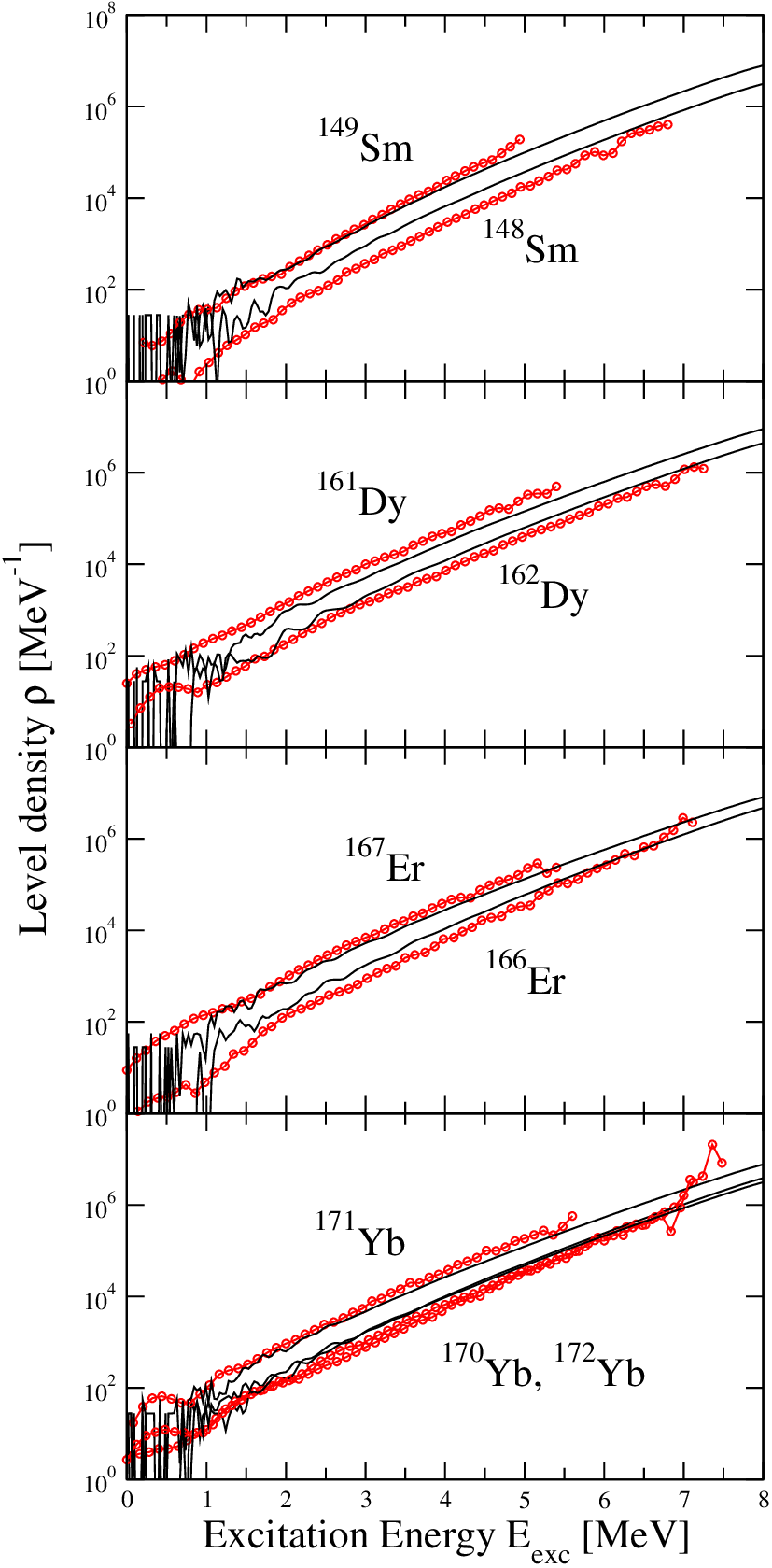}
  \caption{(Color online) Level densities $\rho$ as  functions of excitation energy for
    Sm, Dy, Er and Yb isotopes. The black solid lines show the CFY model
    and
    the red dots show the experimental data
    \cite{OsloSm,OsloDy,OsloEr,OsloYb}. }
  \label{FIG:ALL3}
\end{figure}

\begin{figure}
  \includegraphics[clip=true,width=0.95\columnwidth]{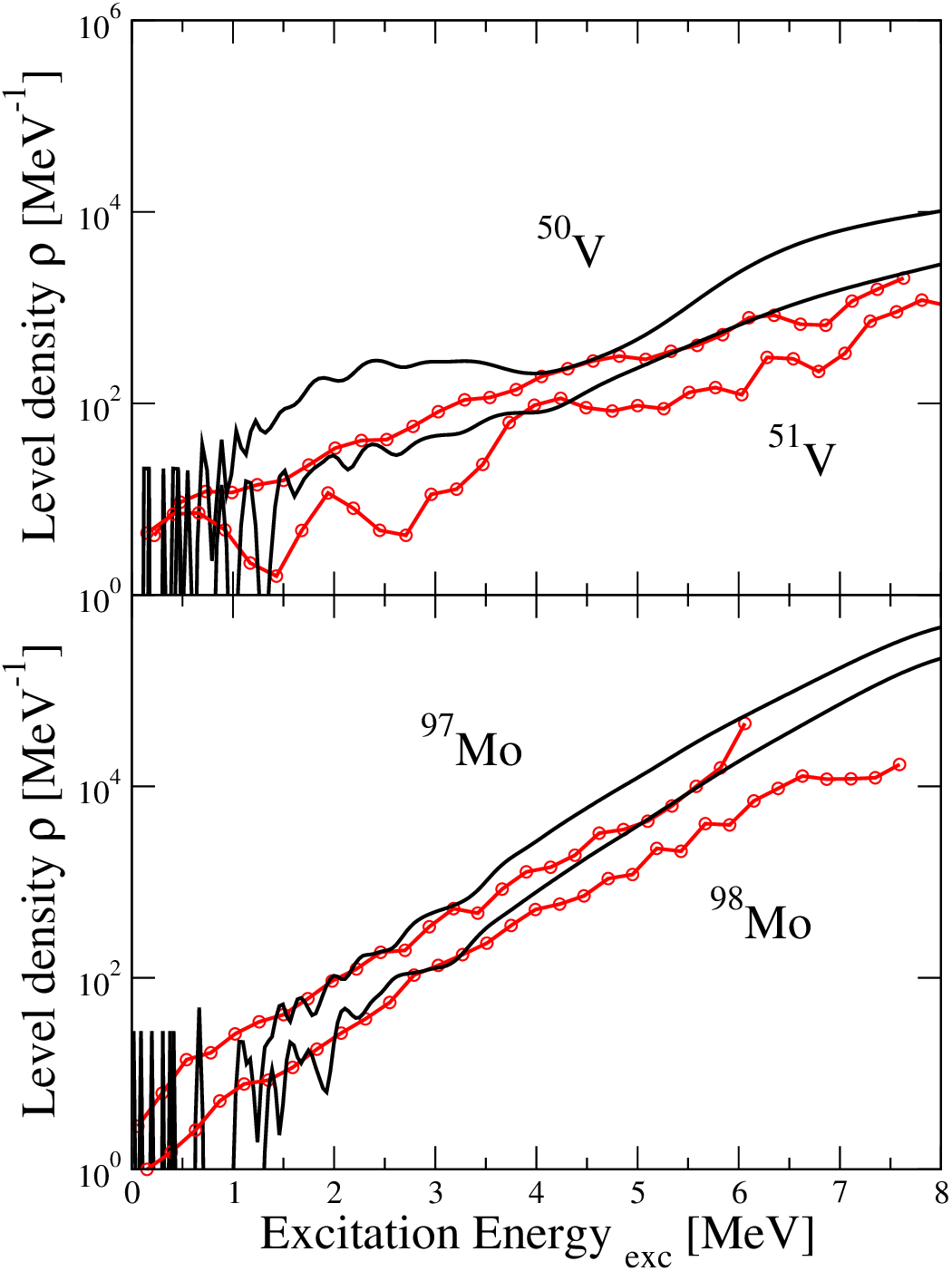}
  \caption{(Color online) Level densities as  functions of excitation energy for
    $^{50,51}$V and $^{97,98}$Mo. The black solid lines show the CFY model and
    the red dots show the experimental data \cite{OsloV,OsloMo}.}
  \label{FIG:V_Mo}
\end{figure}

In Figs.~\ref{FIG:ALL3} and \ref{FIG:V_Mo} level densities obtained
from the CFY model are compared to data for a number of nuclei where
experimental data are available
\cite{OsloV,OsloMo,OsloSm,OsloDy,OsloEr,OsloYb}.

In  Fig.~\ref{FIG:ALL3} we show level densities of the rare-earth nuclei
$^{148,149}$Sm,
$^{161,162}$Dy, $^{166,167}$Er, and
$^{170,171,172}$Yb as functions of excitation
energy.
The data are in general well reproduced by the CFY model, with an
error of less than a factor of 2. The good agreement between the
calculated and experimental slopes  indicates that the
single-particle structure and the moments of inertia in the
rotational bands are sound. However, an observable trend in this
mass region is that the level density for even-even nuclei is
slightly over-estimated and for odd nuclei the level density is
slightly under-estimated. The effective back-shift is to a large
extent controlled by the ground-state pairing gap. By fine-tuning
the pairing gaps it is possible to get an almost perfect agreement
with experiment. However, in this paper no efforts are made to
adjust parameters of the model to fit measured level-density data.
Indeed, the ground-state pairing gaps are given by the mass model,
see Sec.~\ref{sec:pairing}. Neither are any renormalizations of
calculated level densities to fit data performed, as is sometimes
done in other models, see Refs.~\cite{Goriely2006,GorielyOSLO2008}.

Experimental level-density functions are available also for lighter
mass regions, and in Fig.~\ref{FIG:V_Mo} data for V and Mo isotopes
are shown. The overall agreement between the model and these
experimental data is somewhat inferior to what was obtained in the
rare-earth region. For the Mo isotopes in Fig.~\ref{FIG:V_Mo} the
CFY model is roughly a factor of 3 too large for high excitation
energies while for the V isotopes the over-estimation is roughly a
factor of 4. However, these errors are consistent with the overall
results from the neutron separation level spacing, see
Fig.~\ref{FIG:NeutSep}, which have an overall $f_{\rm rms}=4.2$.
Also, the slopes and the detailed structures are not as well
described in these nuclei as in the rare-earths. Especially $^{50}$V
exhibits an oscillatory pattern in the CFY model, which is an effect
of the small ground-state deformation $\varepsilon_2=0.05$
\cite{MollerMass}. A corresponding oscillatory pattern is not as
clearly present in the experimental data. In $^{51}$V the
experimental data shows oscillatory behavior while the CFY model is
more smooth. The ground-state deformation is slightly larger with
$\varepsilon_2=0.083$ \cite{MollerMass}, which in the CFY model
means that since the moment of inertia is larger the rotational
states have a larger influence and smooths the level density.

%-----------------------------------------

\subsection{Low-lying discrete levels}
\label{sec:Accum}

The amount of experimental information on low-lying discrete energy
levels is indeed substantial. In addition to the energy  parity
and angular momentum are often known. In Ref.~\cite{GroundStateSpins} a
global comparison is made between measured and FRLDM calculated
ground-state spins and parities for odd-mass nuclei, and the
agreement was found to be very good. The collection of low-energy
data provides important information for level-density models, since
the accuracy of the model can be studied at the very lowest
energies. Cumulative distributions of known levels can thus be
constructed, and compared to calculations. In
Ref.~\cite{Goriely2006} a selection of 15 different nuclei was made
to represent all different kinds of nuclear aspects, such as
light-heavy, spherical-transitional-deformed, as well as odd-odd,
even-even and odd-even nuclei (also used in
Ref.~\cite{GorielyOSLO2008}). We therefore
compare our model low-lying level densities to
experimental data for the identical
set of 15 nuclei.

\begin{figure}
\centering
\includegraphics[clip=true,width=0.95\columnwidth]{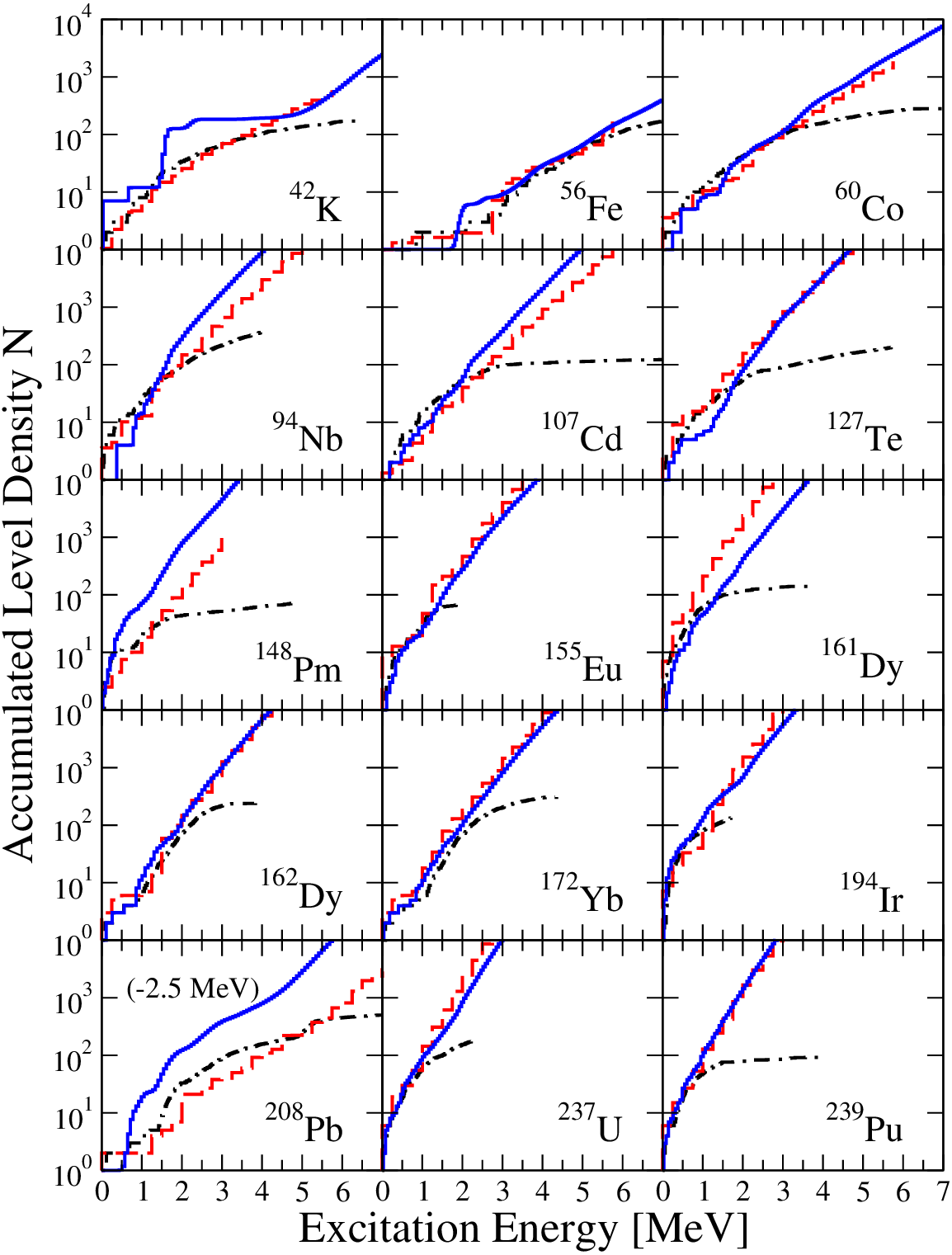}
\caption{(Color online) Accumulated level density as a function of
  excitation energy. The blue solid and red dashed lines show the present CFY
  model and the HFB model \cite{GorielyOSLO2008}. The black
  dash-dotted line shows experimental data collected in Ref.~\cite{GorielyOSLO2008}.}
\label{FIG:Accum}
\end{figure}

The cumulative level density for the 15 nuclei are shown in
Fig.~\ref{FIG:Accum}. The CFY model (solid blue line) is compared to
experimental data (black dash-dotted line) and to the HFB model (red
dashed line) of Ref.~\cite{GorielyOSLO2008}. For a good agreement
with data, the calculated line should follow as precise as possible
the experimental curve at the lowest energies, and should then
smoothly deviate, always larger than data.

In general, the theoretical models seem to give very comparable
results for deformed nuclei while differences are larger for
transitional and spherical nuclei. For the light spherical nuclei,
like $^{42}$K and $^{56}$Fe there are pronounced single-particle gap
effects which cause jumps in the level density. These effects are
smoothly disappearing at higher energies. A quite drastic deviation
between our calculation and data is seen for $^{208}$Pb. The
calculated (cumulative) level density is systematically larger
starting already at about 3 MeV excitation energy (in the figure
seen at 0.5 MeV; the energy scale is shifted by 2.5 MeV for this
nucleus). This is also seen in the total level density at the
neutron separation energy, and in Table~\ref{CumD0table} the ratio
between theoretical and measured level spacings at the neutron
separation energy is listed for the 9 nuclei. For $^{208}$Pb this
ratio is 0.02, i.e. the calculated level density is about 50 times
larger than experiment!

When comparing to the HFB model calculation of the level density
\cite{GorielyOSLO2008} one should note that interpolations between
different deformations are included in this model, while in our
model the ground-state deformation is kept fixed for all excited
states. In particular, this somewhat phenomenological way to treat
deformation changes has strong effects on the results for
$^{127}$Te.

\begin{table}[htb]
  \begin{tabular}{cc|cc|cc}
    \hline
    \hline
     Nucl & $D_{\rm Th}/D_{\rm Exp}$ &Nucl &$D_{\rm Th}/D_{\rm Exp}$ &Nucl & $D_{\rm Th}/D_{\rm Exp}$\\
    \hline
    $^{42}$K & 0.94 & $^{56}$Fe & 0.55 & $^{60}$Co & 0.62\\
    $^{94}$Nb & 2.50 & $^{107}$Cd & 0.77 & $^{127}$Te & 0.26\\
    $^{148}$Pm & 1.72
 & $^{155}$Eu & 2.95 & $^{161}$Dy & 2.61\\
    $^{162}$Dy & 1.27 & $^{172}$Yb & 3.10 & $^{194}$Ir & 4.58\\
    $^{208}$Pb & 0.02 & $^{237}$U & 5.18 & $^{239}$Pu & 3.46\\
    \hline
    \hline
  \end{tabular}
  \caption{Table of neutron separation energy spacings compared to
    experimental data for the nuclei showed in Fig.~\ref{FIG:Accum}.}
  \label{CumD0table}
\end{table}

%-----------------------------------------
\subsection{Parity ratio}
\label{sec:ExpParity}

\begin{figure}
\centering
\subfigure
{
  \includegraphics[clip=true,width=0.95\columnwidth]{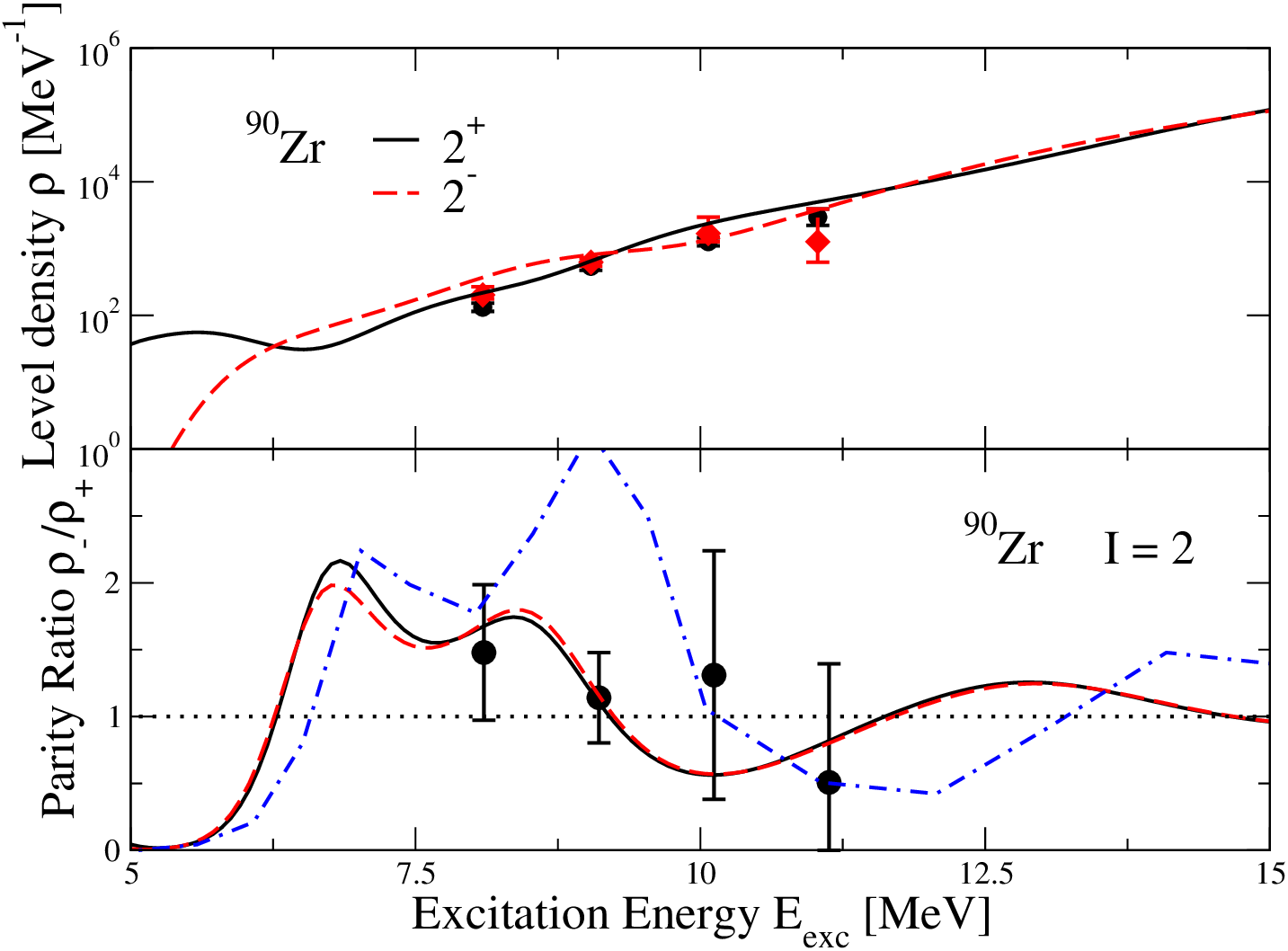}
  \label{FIG:Zr90}
}
\subfigure
{
  \includegraphics[clip=true,width=0.95\columnwidth]{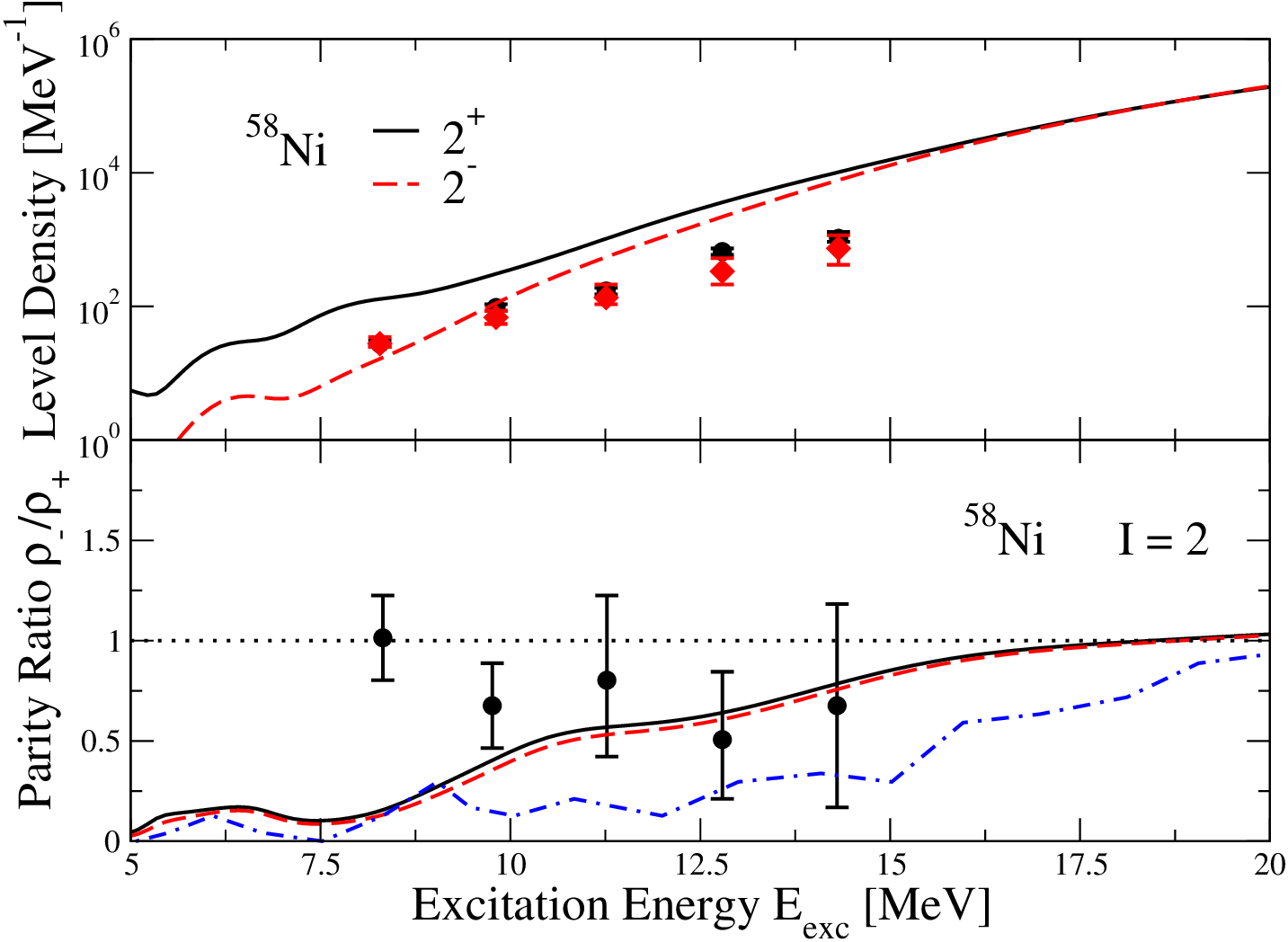}
  \label{FIG:Ni58}
} \caption{(Color online) The top panels show CFY calculated
  components of the $2^+$ (black solid) and $2^-$ (red dashed) level
  densities for  $^{90}$Zr (top part of figure) and $^{58}$Ni (lower
  part of figure), as functions of excitation energy.
  Experimental data from Ref.~\cite{RichterParity} are shown as black dots and
  red diamonds with error-bars for $2^+$ and $2^-$, respectively.
  The lower panels show for each nucleus the parity ratio
  versus excitation energy. The solid
  black and dashed red lines show the CFY model results for the total level density
  and the I=2 component, respectively. The blue dot-dashed lines show the Skyrme-HFB model of
  Ref.~\cite{Goriely2006}. Experimental data from Ref.~\cite{RichterParity} are shown with error-bars.}

\end{figure}

The parity ratio has been measured experimentally by Kalmykov et.al.
\cite{RichterParity} for the two spherical nuclei $^{58}$Ni and
$^{90}$Zr. For these nuclei the
 $I=2$ angular-momentum component of the level density is measured and
separated into parity components. In the top panel of
Fig.~\ref{FIG:Zr90} the level-density component $\rho(E,I=2,\pi=\pm
1)$ for $^{90}$Zr is compared to CFY model calculations. The level
densities from the CFY model are in good agreement with experimental
data, and in the lower panel of Fig.~\ref{FIG:Zr90} the parity ratio
is shown. Predictions for the total level density as well as the
$I=2$ component are shown as black solid and red dashed lines,
respectively. The difference between the parity ratio for the total
level density and the $I=2$ component decreases with excitation
energy, because the spin cutoff model becomes more realistic when
the excitation energy increases, see Sec.~\ref{sec:spindistr}. For
the parity ratio the CFY model result is within the experimental
error-bars for all excitation energies and seems to show a similar
pattern as the experiments:  a high parity ratio (excess of negative
parity states) at 8 MeV and a low ratio (excess of positive parity
states) at 11 MeV excitation energy. The blue dot-dashed line shows
the Skyrme-HFB model of Ref.~\cite{Goriely2006}. It is also within
the experimental error-bars for all excitation energies except at 9
MeV where the HFB model gives a large positive ratio ($\sim 3$)
while the experiments and the CFY model are close to unity.

The $I=2$ component of the level density in $^{58}$Ni is shown in
the top panel of Fig.~\ref{FIG:Ni58}. The agreement between
experimental data and the CFY model is somewhat inferior to what we
obtained for $^{90}$Zr.  For both parities the level density seems
to increase faster in the CFY model than what is seen  in the
experimental data. At 14 MeV excitation energy the model
overestimates the level density by roughly a factor of 6. The parity
ratio is shown in the lower panel of Fig.~\ref{FIG:Ni58}. The
experimental parity ratio is close to unity at 8 MeV and then
decreases with increasing excitation energy. The CFY model shows a
different pattern. It gives  a very low parity ratio at low
excitation energies with an almost monotonic increase with
increasing excitation energy. The ratio only becomes close to unity
at 20 MeV\@. The Skyrme-HFB model of Ref.~\cite{Goriely2006} shows a
similar pattern as the CFY model but with an even lower ratio for
excitation energies below 20 MeV.

\begin{figure}
  \includegraphics[clip=true,width=0.95\columnwidth]{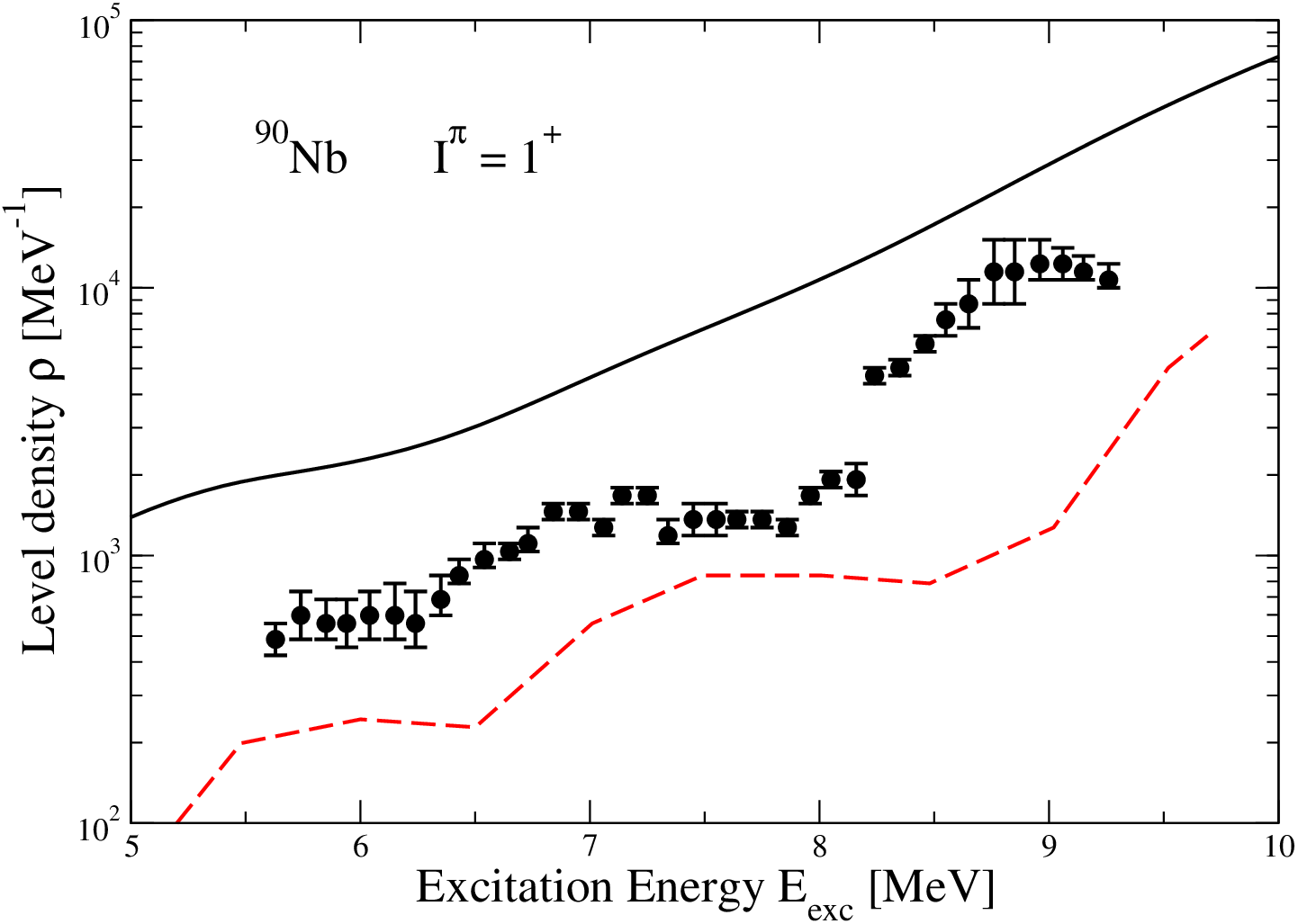}
  \caption{(Color online) Level density of the $1^+$
    component of the level density as a function of excitation energy for
    $^{90}$Nb. The
    black solid and red dashed lines show the CFY model and Skyrme-HFB
    model of Ref.~\cite{Goriely2006}, respectively. Data
    from Ref.~\cite{RichterParity} are shown as black dots with error-bars.}
  \label{FIG:Nb90}
\end{figure}

The $1^+$ component of the level density of $^{90}$Nb has also been
measured in Ref.~\cite{RichterParity}. Fig.~\ref{FIG:Nb90} shows the
experimental data compared to the CFY and Skyrme-HFB models. In the
experimental data there is a clear oscillating structure. The CFY
model (black solid line) over-estimates the level density and the
oscillating structure is much less pronounced. In the Skyrme-HFB
model (red dashed line) there are long-range oscillations similar to
what is seen in the experimental data, but the energy separation
between consecutive minima is larger. The level density is
under-estimated in the Skyrme-HFB model, whereas  the CFY model
over-estimates the level density by a similar factor.

%%%%%%%%%%%%%%%%%%%%%%%%%%%%%%%%%%%%%%%%%%%%%%%%%%%%%%%%%%%%%%%

\section{Comparison with other models}
\label{sec:othermodels}

\begin{table}
  \begin{tabular}{lll}
    \hline
    \hline
    Statistical Models & $f_{\rm rms}$ & Ref.\\
    \hline
    Back-shifted Fermi Model &1.71&\cite{Goriely2008}\\
    Const. Temp. Model &1.77&\cite{Goriely2008}\\
    Back-shifted Fermi + Const. Temp. Model &1.7&\cite{Rauscher}\\
    Generalized Superfluid Model &1.94&\cite{Goriely2008}\\
    \hline
    \hline
    &&\\
    \hline
    \hline
    Combinatorial Models & $f_{\rm rms}$ & Ref.\\
    \hline
    Skyrme-HFB &2.35&\cite{Goriely2008}\\
    Gogny-HFB & 4.55 & \cite{HilaireGirod}\\
    CFY & 4.2 & Present\\
    \hline
    \hline
  \end{tabular}
  \caption{Table of rms-factors $f_{\rm rms}$ for statistical and
    combinatorial models for neutron separation energy level spacings
    \cite{Goriely2008,Rauscher,HilaireGirod}.}
  \label{RMStable}
\end{table}

The CFY model is here compared to other statistical and
combinatorial NLD models that provide neutron resonance level
spacings. In general, the statistical models listed in
Table~\ref{RMStable} have  an rms deviation just below 2. This low
rms deviation is obtained because several parameters in the
level-density formulas are directly adjusted to the neutron
separation-energy level spacings and low-lying discrete energy
levels.

The back-shifted Fermi model of Ref.~\cite{Goriely2008} is based on a
simple Fermi-gas formula whereas the Constant Temperature model is based on
the approach of Gilbert and Cameron \cite{GilbertCameron}. Both models contain
explicit enhancement factors for rotations and vibrations. The model by
Rauscher, Thielemann and Kratz \cite{Rauscher} is also based on the approach
of Gilbert and Cameron. In contrast to the first two models this model has no
explicit enhancement factors
for rotations and vibrations. It accounts for  shell effects and thermal damping in
terms of an effective level-density parameter. The model uses the microscopic energy
corrections of Ref.~\cite{MollerMass} together with three free parameters in
the fitting procedure. The Generalized Superfluid Model is similar to the
models mentioned above. In addition it takes into account how pairing
evolves  with increasing excitation energy. It also incorporates explicit
rotational and vibrational enhancements \cite{Goriely2008}.

\begin{figure}
  \includegraphics[clip=true,width=0.95\columnwidth]{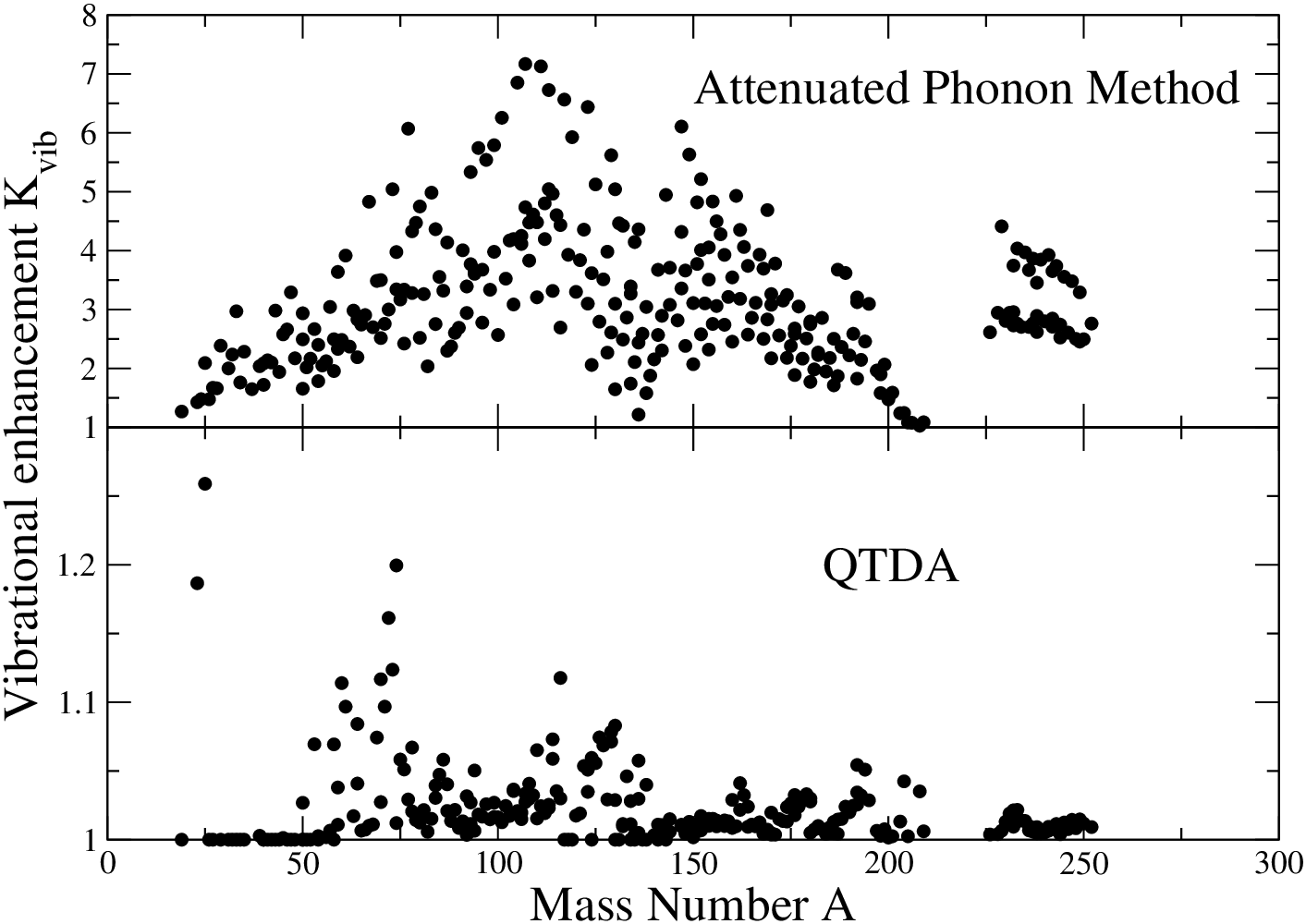}
  \caption{Vibrational enhancements at the neutron separation energy in the
    attenuated phonon method (top panel)
    and the QTDA method (bottom panel) versus of mass number A.}
  \label{FIG:NeutSepVib}
\end{figure}

Predictions by  statistical models in regions outside the fitting
region probably give much less accurate results than in the
adjustment region. On the other hand, since  the combinatorial
models are all based on calculated  single-particle spectra they
could in principle  have better predictive power, in particular in
regions where the single-particle model is sound. In contrast, they
are not equally flexible in terms of  parameter fits to the level
spacings database. Therefore, the rms deviation factor of
combinatorial models with respect to known data  is larger than for
the statistical models. The $f_{\rm rms}=4.2$ for the CFY model is
about a factor of 2 larger than in statistical models. However, the
latter  result is  obtained with \emph{no} parameters specifically
fitted to the level-density data. The mean deviation factor $m=1.1$
indicates that the level density is on average reasonable. However,
there seems to be a drift with mass number, as seen in
Fig.~\ref{FIG:NeutSep}, present in the CFY model which is not seen
in the HFB models of Refs.~\cite{HilaireGirod,Goriely2008}.

In the lower part of Table~\ref{RMStable} the CFY model is compared
to two other large-scale combinatorial NLD models based on the HFB
method. The model of Ref.~\cite{HilaireGirod} is based on the Gogny
D1S interaction for the mean-field and incorporates combinatorial
rotations (similar to Sec.~\ref{sec:ROT} but no pairing dependence
of the moment of inertia), and the attenuated phonon method is
applied to account for vibrational states. Pairing is included
explicitly for the ground-state, and excited states are back-shifted
by an energy-dependent gap procedure. The model gives $f_{\rm
  rms}=4.55$ for the subset of even-even axially deformed nuclei. The error
is slightly larger than the CFY model.

For the Skyrme-HFB NLD model the
deviation factor is $f_{\rm rms}=2.35$ \cite{Goriely2008}, which is
$\sim 35\%$ larger than the statistical models and 44\% smaller
that in the CFY model. This model is based on a Skyrme-HFB
mean-field together with combinatorial rotations and the attenuated phonon
method for modeling vibrational states. In addition a
phenomenological deformation change from deformed to spherical shape at some
specified excitation energy is incorporated.

The attenuated phonon method is a phenomenological way to model nuclear
vibrations, see eg. Refs.~\cite{GSM,Goriely2006}. It assumes that the
quadrupole and octupole phonon states in nuclei can be modeled by a gas of
non-interacting bosons. The vibrational excitation energies are taken from
systematics of the
lowest non-rotational $2^+$ and $3^-$ states. The model is formulated as a
multiplicative factor $K_{\rm vibr} = \exp \left[ \delta S -\delta U/T\right]$
which enhances the level density. $T$ is the nuclear temperature
and the $\delta S$ and $\delta U$ are the entropy and internal energy change
induced by the bosons. The occupation probabilities of the bosons are
described by damped Bose statistics, where the damping sets in at considerably
higher energies than the neutron separation energy. At excitation energies
close to or below the neutron separation energy the damping has negligible
effect which implies that the phonons are allowed to be repeated several
times.

The vibrational enhancement at the neutron separation energy is
quite small in the QTDA method as compared to the attenuated phonon
method, see Fig.~\ref{FIG:NeutSepVib}. The QTDA gives an enhancement
of the order of a few percent compared to up to a factor 7 in the
attenuated phonon method. The largest vibrational enhancements are
obtained for the $A$=75 region and for Cd isotopes in the $A$=115
region, which is consistent with the strong (quadrupole) vibrational
character of nuclei in these regions.

The QTDA phonons are microscopically built from quasi-particle
excitations $E_{\mu}^{\rm qp,i}$ with energies not much different
than the QTDA phonon energies $(\hbar \omega)_j^i$. The way to fully
account for double-counting of phonon states
(Eq.~\ref{levdenschange}), thus implies a small vibrational
enhancement since the existence of a few low-lying collective
phonons will not impact the level density if all other phonons are
very non-collective. The non-collective character of most of the
phonons also implies that they cannot be repeated. Phonons in the
attenuated phonon model are not described microscopically, and
double-counting of states may clearly appear. In addition, it
assumes that each phonon can be repeated several times.

%%%%%%%%%%%%%%%%%%%%%%%%%%%%%%%%%%%%%%%%%%%%%%%%%%%%%%%%%%%%%%%

\section{Summary}
\label{sec:summary}

A combinatorial model for the nuclear level density is presented.
The model is based on the folded-Yukawa single-particle model with
ground-state deformations and parameters from
Ref.~\cite{MollerMass}. The model is used to calculate the neutron
resonance level spacings, yielding an rms deviation of $f_{\rm
rms}=4.2$. It also compares favorably with experimental level
density data versus excitation energy for several nuclei in the
rare-earth region as measured by the Oslo method, as well as with
the cumulative level density extracted from low-energy spectra.

The role of collective enhancements has been investigated in detail.
Pairing is incorporated for each individual many-body configuration,
and the distribution of the pairing gaps is investigated. No sharp
pairing phase transition is observed. Instead, even at the highest
excitation energy considered, a non-negligible fraction of the
states have a considerable pairing gap.

Rotational states are included combinatorially by a simple rotor model with a
moment of inertia dependent on the deformation and pairing gap. Vibrational
states are included using a Quasi-particle-Tamm-Dancoff Approximation. It is
found that the vibrational enhancement in the QTDA model is very small, on the
order of a few percent at the neutron separation energy.

The parity distribution in the CFY model shows large oscillatory
patterns for nuclei which have large gaps in the single-particle
spectrum, separating shells with different parities. This is often
the case for nuclei with small deformations. The patterns are quite
different from the smooth pattern of the statistical parity
distribution model of Ref.~\cite{AlhassidParity}. The CFY model is
compared to other models and to experimental data when available.

The angular-momentum distribution in the CFY model is compared to the
spin cutoff model in Sec.~\ref{sec:spindistr}. It is found that the
Gaussian envelope of the spin cutoff model is in good agreement with the CFY
model for high excitation energies.

\begin{acknowledgments}
We acknowledge discussions with R. Capote, M. Guttormsen, K.-L.
Kratz, T. Rauscher, A. Richter and F.-K. Thielemann. S. Hilaire and
S. Goriely are acknowledged for providing data underlying Fig.14.

H. U. is grateful for the hospitality of the Los Alamos National
Laboratory during several visits. S. \AA . and H. U. thank the
Swedish national research council (VR) for support. This work was
supported by travel grants for P. M.\ to JUSTIPEN (Japan-U. S.
Theory Institute for Physics with Exotic Nuclei) under grant number
DE-FG02-06ER41407 (U. Tennessee). This work was partially carried
out under the auspices of the National Nuclear Security
Administration of the U. S. Department of Energy at Los Alamos
National Laboratory under Contract No.\ DE-AC52-06NA25396 and DE-FC02-07ER41457.

\end{acknowledgments}

%%%%%%%%%%%%%%%%%%%%%%%%%%%%%%%%%%%%%%%%%%%%%%%%%%%%%%%%%%%%%%%

\end{document}